\documentclass[lettersize,journal]{IEEEtran}
\usepackage{amsmath,amsfonts,amssymb}
\usepackage{algorithmic,algorithm}
\usepackage{array}
\usepackage[caption=false,font=normalsize,labelfont=sf,textfont=sf]{subfig}
\usepackage{textcomp}
\usepackage{stfloats}
\usepackage{url}
\usepackage{verbatim}
\usepackage{graphicx}
\usepackage{cite}
\usepackage{epstopdf}
\usepackage{color}
\usepackage{amsthm}
\usepackage{makecell}
\usepackage{color}
\newtheoremstyle{examplestyle}
{ }
{ }
{\normalfont}
{\parindent}
{\itshape}
{:}
{ }
{}
\theoremstyle{examplestyle}
\newtheorem{prn}{Remark}
\newtheorem{pry}{Property}
\newtheorem*{prf}{Proof}
\usepackage{fancyhdr}
\usepackage{kantlipsum}
\fancypagestyle{plain}{
	\fancyhf{}
	\fancyhead[C]{Conference on \LaTeX}     
	\fancyfoot[L]{This is a notice}

}
\usepackage{eso-pic}

\begin{document}
	\AddToShipoutPictureBG*{%
		\AtPageUpperLeft{%
			\setlength\unitlength{1in}%
			\hspace*{\dimexpr0.5\paperwidth\relax}
			\makebox(0,-0.75)[c]{\small This article has been accepted for publication in IEEE Journal on Selected Areas in Communications.}%
			\makebox(0,-1.0)[c]{\small This is the author's version which has not been fully edited and content may change prior to final publication.}%
	}}
	
	\title{Joint Uplink and Downlink Resource Allocation Towards Energy-efficient Transmission for URLLC}
	
	\author{Kang Li,~Pengcheng Zhu,~\IEEEmembership{Member,~IEEE},~Yan Wang,\\~Fu-Chun Zheng,~\IEEEmembership{Senior Member,~IEEE},~and Xiaohu You,~\IEEEmembership{Fellow,~IEEE}
		\thanks{$\copyright$ 20XX IEEE. Personal use of this material is permitted. Permission from IEEE must be obtained for all other uses, in any current or future media, including reprinting/republishing this material for advertising or promotional purposes, creating new collective works, for resale or redistribution to servers or lists, or reuse of any copyrighted component of this work in other works.}
		\thanks{This work was supported in part by the National Key Research and Development Program of China under Grant 2021YFB2900300, the National Natural Science Foundation of China under Grant 62171126, and the Shenzhen Science and Technology Program under Grants KQTD20190929172545139 and ZDSYS20210623091808025. (Corresponding author: Pengcheng Zhu; Fu-Chun Zheng.)}
		\thanks{Kang Li and Yan Wang are with the National Mobile Communications Research Laboratory, Southeast University, Nanjing 210096, China. (e-mail: ricoh@seu.edu.cn; yanwang@seu.edu.cn).}
		\thanks{Pengcheng Zhu and Xiaohu You are with the National Mobile Communications Research Laboratory, Southeast University, Nanjing 210096, China, and also with the Purple Mountain Laboratories, Nanjing 211111, China. (e-mail: p.zhu@seu.edu.cn; xhyu@seu.edu.cn).}
		\thanks{Fu-Chun Zheng is with the School of Electronic and Information Engineering, Harbin Institute of Technology (Shenzhen), Shenzhen 518055, China. (e-mail: fzheng@ieee.org).}}
	
	
	\maketitle
	\begin{abstract}
		Ultra-reliable and low-latency communications (URLLC) is firstly proposed in 5G networks, and expected to support applications with the most stringent quality-of-service (QoS). However, since the wireless channels vary dynamically, the transmit power for ensuring the QoS requirements of URLLC may be very high, which conflicts with the power limitation of a real system. To fulfill the successful URLLC transmission with finite transmit power, we propose an energy-efficient packet delivery mechanism incorparated with frequency-hopping and proactive dropping in this paper. To reduce uplink outage probability, frequency-hopping provides more chances for transmission so that the failure hardly occurs. To avoid downlink outage from queue clearing, proactive dropping controls overall reliability by introducing an extra error component. With the proposed packet delivery mechanism, we jointly optimize bandwidth allocation and power control of uplink and downlink, antenna configuration, and subchannel assignment to minimize the average total power under the constraint of URLLC transmission requirements. Via theoretical analysis (e.g., the convexity with respect to bandwidth, the independence of bandwidth allocation, the convexity of antenna configuration with inactive constraints), the simplication of finding the global optimal solution for resource allocation is addressed. A three-step method is then proposed to find the optimal solution for resource allocation. Simulation results validate the analysis and show the performance gain by optimizing resource allocation with the proposed packet delivery mechanism.
	\end{abstract}
	
	\section{Introduction}
	\IEEEPARstart{U}{ltra-reliable} and low-latency communications (URLLC) has been considered as the key enabler to support various mission-critical applications under the fifth generation (5G) and beyond networks\cite{You2021SciChinaInfSci}, such as autonomous driving, factory automation and remote healthcare\cite{Bennis2018jproc1834}. According to the 3GPP (3rd Generation Partnership Project) standard, a general URLLC requirement is the target reliability of 99.999\% for transmitting a packet of 32 bytes within a user plane latency of 1 ms\cite{TR38.913}. The reliability is defined as the percentage of packets that are correctly received. Depending on the applications, the end-to-end (E2E) latency within several to tens of milliseconds includes but is not limited to uplink (UL) and downlink (DL) transmission delay, coding and processing delay as well as queueing delay\cite{Chen2018mcom119}. Such stringent requirements have posed many challenges on system design from physical layer to network layer.
	
	In the long-term evolution (LTE) networks, the transmission time interval (TTI) which is the minimum time granularity for scheduling is set as 1 ms\cite{Capozzi2013surv678}, and hence unable to satisfy the E2E latency requirement of URLLC. To reduce latency, one efficient way is to employ the short frame structure on physical layer\cite{TR38.802}. By increasing the subcarrier spacing (SCS) and/or reducing the number of symbols, the TTI could be shortened to no more than 0.2-0.25 ms\cite{TR38.912}. However, with the short frame structure and small packet size in URLLC, the blocklength of channel coding is short, such that the transmission error cannot be ignored\cite{Shannon1948}. Furthermore, due to the features of URLLC, the transmit power for ensuring the target reliability may need to be very high when the channel stays in deep fading\cite{She2018twc127}. This brings challenges on link layer transmission for a real system, since the maximum transmit power of a device or BS is limited. As a result, other supporting technologies in conjunction with their resource allocation are needed for URLLC to guarantee the target reliability and other quality-of-service (QoS) metrics.
	
	\subsection{Related Work}
	When it comes to the traditional services where the blocklength is sufficiently long (e.g., 1500 bytes in human-to-human communications\cite{TR36.814}), Shannon's capacity has been widely used to characterize the maximum achievable rate with the decoding error arbitrarily close to zero\cite{Shannon1948}. In URLLC, to satisfy the low-latency requirement (say 1 ms in E2E latency), the packets of small size (say 20 bytes\cite{TR38.913}) using short frame structure must be transmitted. As a result, the blocklength of channel coding becomes short, and the decoding error can not be ignored. Therefore, if Shannon’s capacity is used in optimizing resource allocation for URLLC, the latency and reliability will be underestimated\cite{Xu2016twc5527}. In the seminal work, Polyanskiy et al.\cite{Polyanskiy2010tit2307} derived a normal approximation on the maximum achievable rate with short blocklength channel codes over additive white Gaussian noise (AWGN) channel. The results indicate that, at the expense of achievable rate reduction, a low transmission error probability in the short blocklength region can be guaranteed. The study was extended to mutiple antennas system over quasi-static channel in \cite{Polyanskiy2014tit4232}, for cases with and without channel state information (CSI) at the transmitter and/or the receiver. Popovski et al. \cite{Popovski2016jproc1711} further applied the maximum achievable rates obtained in \cite{Polyanskiy2010tit2307} and \cite{Polyanskiy2014tit4232} to the two-way channel, the downlink broadcast channel, and the uplink random access channel. However, even with the simplified approximation, the expression of maximum achievable rate is still neither convex nor concave with respect to bandwidth and power\cite{She2018tcomm2266,She2019twc402}. As a result, a globally optimal radio resource allocation for URLLC with such an achievable rate is still hard to obtain.
	
	Based on the theoretical principles that govern the transmission of short packets, optimizing resource allocation towards improving spectrum efficiency (SE) or energy efficiency (EE) for URLLC has gained attention\cite{Xu2016twc5527,She2018tcomm2266,She2019twc402,Liu2022lwc}. However, how to achieve the optimization goal without sacrificing the QoS reuqirements of URLLC is a critical problem. In URLLC, the required latency and bandwidth do not exceed the channel coherence time and coherence bandwidth\cite{Polyanskiy2014tit4232}. As a result, the target reliability of URLLC is hard to guarantee when the channel stays in deep fading. To support high reliability over the fading channels, retransmission and diversity as key techniques have been exploited in the existing literature. Since the retransmission procedure of hybrid automatic repeat request (HARQ) introduces additional latency\cite{Pedersen2017mwc154}, other retransmission schemes are needed for URLLC. Focusing on the two classes of packets in Industrial Internet of Things (IIoT), two retransmission schemes, namely, retransmission with individual reservation and retransmission with contention-based reservation jointly with packet repetitions, are proposed to meet the URLLC requirements with low resource consumption in \cite{Elayoubi2019jsac896}. Unfortunately, retransmission schemes can hardly improve the success probability when the channels stay in deep fading within the coherence time \cite{She2018twc127}. As a result, diversity technique together with resource allocation is promising to ensure the target reliability of URLLC. The authors in \cite{She2016GLOCOMW} studied the problem of optimal bandwidth allocation that ensures QoS metrics with frequency diversity in UL, and the authors in \cite{She2017GLOCOMW} investigated optimal transmit power allocation under QoS constraints with multi-user diversity in DL. As regards spatial diversity, \cite{Johansson2015ICCW} indicates that the required transmit power to ensure reliability can be reduced when the number of antennas at a base station (BS) increases. In addition, She et al.\cite{She2018twc127} proposed an original proactive dropping scheme to ensure the overall reliability of URLLC by controlling the packet dropping probability with resource allocation. However, these works only allocate either UL or DL resources, and generally each above-mentioned technique improves the reliability at the expense of paying another price, e.g., frequency diversity needs redundant bandwidth, while proactive dropping introduces an extra error component. For URLLC, the required transmit power that ensures QoS metrics may need to be very high when the channel stays in deep fading\cite{She2018twc127}, but this sounds conflicted with the constraint of maximum transmit power of a device or BS. Therefore, by jointly allocating UL and DL resources under new transmission mechanism, this paper focuses on the optimization of power consumption so as to improve EE, while satisfying the QoS requirements of URLLC and the constraint of maximum transmit power.
	
	\subsection{Contributions}
	In this paper, we study how to obatin the global optimal resource allocation for the proposed packet delivery mechanism, to minimize the average total power under the constraint of URLLC transmission requirements in local communications scenario. Although technical challenges on URLLC system design exist at different layers, we foucus on link layer design here to achieve the QoS requirements of URLLC from a joint UL and DL transmission aspect. The major contributions of this work are summarized as follows:
	\begin{itemize}
			\item We propose a packet delivery mechanism for URLLC in machine-type communications (MTC). To reduce the outage probability and increase the number of participating sensors in UL, we consider frequency-hopping to provide more chances for transmission with the introduced subchannels, so that the outage due to channel deep fading hardly occurs. To avoid the downlink outage from queue clearing and increase the number of participating users in DL, we consider proactive dropping to ensure the overall reliability under channel deep fading by discarding certain packets.
			
			\item We prove that under the proposed delivery mechanism, a low packet loss probability can be ensured with finite power by exploiting sufficient spatial diversity gain. For the case where spatial diversity gain is insufficient, there has to be a minimum number of antennas to guarantee the target reliability of URLLC with finite power. A binary method is proposed to determine the required minimum number of antennas. Furthermore, simulation results show that the subchannels introduced by frequency-hopping can help reduce the required minimum number of antennas.
			
			\item We jointly optimize resource allocation, including the bandwidth, the threshold of transmit power, the number of antennas and subchannels, to minimize the average total power under the constraints of maximum transmit power and QoS metrics of URLLC. Via theoretical analysis (e.g., the convexity with respect to bandwidth, the independence of bandwidth allocation, the convexity of antenna configuration with inactive constraints), the simplication of finding the global optimal solution is addressed. A three-step method is then proposed, where the bandwidth allocation, antenna configuration and subchannel assignment are optimized in turn. Simulation results show that the optimal solution under the proposed delivery mechanism can save more power than other allocation strategies, and thus enhance EE compared with other transmission schemes. 
	\end{itemize}

	The rest of this paper is organized as follows: Section \ref{II} describes an example system model and the QoS requirements of URLLC. Section \ref{III} reviews the possible transmission schemes, and proposes a packet delivery mechanism for joint UL and DL transmission. In section \ref{IV}, an optimization problem that minimizes the average total power is formulated. Section \ref{V} illustrates how to obtain optimal resource allocation. Simulation results are provided in Section \ref{VI}. In the end, Section \ref{VII} concludes this work.
	\section{System Model}\label{II}
	\begin{figure}[!t]
		\centering
		\includegraphics[width=3.5in]{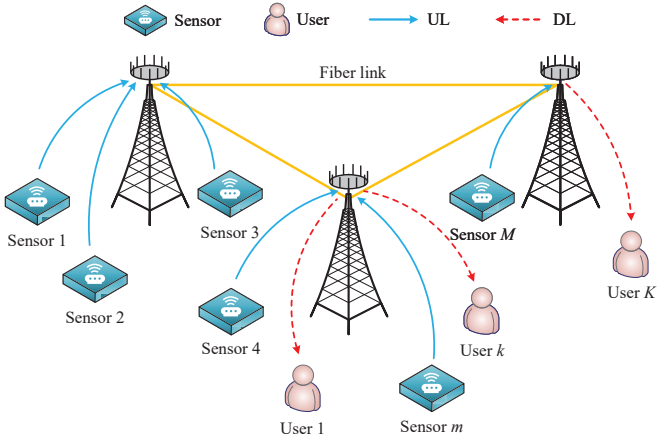}
		\caption{Cellular network in local communications scenario.}
		\label{fig:system}
	\end{figure}
	We consider a cellular network in local communications scenario,\footnote{Depending on mission-critical services, there are three typical communication scenarios for URLLC: local communications, mobile edge computing, and wide-area large-scale networks\cite{Feng2019mvt94}. Each of them has different network architectures, and hence the factors and delay components that lead to packet loss are not the same. In this paper, we focus on how to guarantee the QoS metrics of URLLC in local area communications scenario as stated in \cite{TS22.368}.} whose covergae is less than a few kilometers. The cellular network is composed of one or several adjacent BSs which are interconnected with one-hop fiber links, where each BS is euippped with $N_\mathrm{t}$ antennas. In the coverage, there are $M+K$ single-antenna devices classified into two groups in terms of the functionality: sensors that only upload packets and users that only download packets which are uploaded by the sensors. As illustrated in Fig. \ref{fig:system}, when a sensor and its intended user are not served by the same BS, the packet uploaded by the sensor will go through the folllowing procedure: At first, the packet is forwarded to the BS connected with the intended user via fiber links. Then, it waits in the associated queue of the BS. In the end, the BS transmits the packet to the intended user.
	
	Such a system model can be applied in analyzing the QoS requirements of MTC services (e.g., autonomous vehicles and factory automation) which are representative for URLLC\cite{Bennis2018jproc1834}\footnote{Device-to-device (D2D) communication mode is also applicable, but the performance on reliability and available range is inferior to the cellular mode\cite{She2018tcomm5482}. How to better use D2D mode for URLLC deserves further study but will not be addressed in this work.}. We take a frequency division duplex (FDD) system as an example, where the total bandwidth is shared by UL and DL transmissions.  Orthogonal frequency division multiple access (OFDMA) and frequency reuse are adopted to avoid intra-cell interference and inter-cell interference, respectively.
	
	\subsection{Channel Model}
	To accord with the channel characteristics of URLLC, we consider quasi-static flat fading channel\cite{Polyanskiy2014tit4232}, where the channel remains constant within the channel coherence time, and the bandwidth allocated to device is smaller than the channel coherence bandwidth. This assumption is easy to get satisfied in URLLC. On the one hand, the required bandwidth to transmit a small size packet does not exceed the channel coherence bandwidth, which is around 0.5 MHz at the maximum delay spread of 1 $\mu$s in MTC\cite{Tse2005}. On the other hand, the latency requirement is shorter than the channel coherence time, such as the coherence time of 5-10 ms when the velocity of device is less than 120 km/h\cite{She2017SciChinaInfSci}. For a single-input-multiple-output (SIMO) or multiple-input-single-output (MISO) system subject to quasi-static flat fading channel, the maximum achievable rate with finite blocklength coding can be approximated as\cite{Popovski2016jproc1711}
	\begin{equation}
		\begin{aligned}
			R \approx\frac{B }{\ln2}\left[\ln\left(1+\frac{\mu g P}{\phi N_0 B }\right)-\sqrt{\frac{V}{\tau B }}f_Q^{-1}(\varepsilon^{\mathrm{c}})\right]\left(\text{bits/s}\right).\label{eq:rate}
		\end{aligned}
	\end{equation}
	Note that (\ref{eq:rate}) is obtained for interference-free systems,\footnote{The maximum achievable rate with finite blocklength in interference channels is unavailable in the literature until now. Therefore, strong interference should be eliminated to ensure reliability in URLLC.} where $B$ is the bandwidth, $P$ is the transmit power, $\phi >1$ is the SNR loss due to imperfect CSI,\footnote{The impact of imperfect CSI brought by channel estimation errors can be equivalent to an SNR loss on data rate, which depends on the velocity of devices\cite{Liu2015tvt2846}. For devices with slow and medium velocity, $\phi$ is close to 1\cite{She2018tcomm2266}.} $\mu$ is the large-scale channel gain depending on path loss and shadowing, $g$ is the small-scale channel gain affected by multi-path fading, $N_0$ is the single-side noise spectral density, $\tau$ is the data transmission duration, $\varepsilon^{\mathrm{c}}$ is the decoding error probability, $f_Q^{-1}(x)$ is the inverse of the Gaussian Q-function, and $V=1-\left(1+\frac{\mu g P}{\phi N_{0} B}\right)^{-2}$ is the channel dispersion\cite{Popovski2016jproc1711}.
	
	As shown in \cite{Polyanskiy2010tit2307} and \cite{Polyanskiy2014tit4232}, the approximation in (\ref{eq:rate}) is very accurate at $\varepsilon^{\mathrm{c}}\in [10^{-3},10^{-6}]$. As validated in \cite{Schiessl2015MSWiM}, the channel dispersion is almost 1 when the received SNR is higher than 5 dB. Now that the typical SNR is around 10 dB at the edge of a cell\cite{Tse2005}, $V\approx1$ is applied so that the lower bound of (\ref{eq:rate}) can be obtained. For the resource allocation, if the lower bound is used for optimization, the QoS metrics when $V < 1$ in low SNR regime can be naturally satisfied.
	
	Denote the channel vector as $\boldsymbol{h}$ with independent identically distributed (i.i.d.), zero mean, unit variance and complex circularly symmetric Gaussian random elements (i.e., zero-mean spatially white). For SIMO or MISO system, the small-scale channel gain is $g=\boldsymbol{h}^{H} \boldsymbol{h}$ whose probability density function (PDF) under Rayleigh fading is given by $f_{N_{\mathrm{t}}}(g)=\frac{g^{N_{\mathrm{t}}-1}}{\left(N_{\mathrm{t}}-1\right) !} e^{-g}$\cite{Paulraj2008}, where $\boldsymbol{h}\in \mathbb{C}^{N_{\mathrm{t}} \times 1}$ and $(\cdot)^{H}$ denotes conjugate transpose. To further improve SNR in SIMO or MISO system, maximum ratio transmission/combining (MRT/MRC) is adpoted to maximize the signal gain.
	
	\subsection{Traffic Model}
	In reality, not all sensors are simultaneously activated to transmit packets. As a result, whether each sensor has a packet to transmit in UL could be modelled as a Bernouli process (i.e., active with probability $\kappa$ or silent with probability $1-\kappa$ in each frame). Let $\mathcal{A}_{k}$ denote the set of active sensors that transmit packets to the $k$-th user. Assume that each packet arrival process is i.i.d.. Then, the aggregation of these arrival processes from $\left|\mathcal{A}_{k}\right|$ sensors is a Poisson process, where the average arrival rate is $\lambda_{k}\triangleq\left|\mathcal{A}_{k}\right| \kappa$ packets/frame,\footnote{The average arrival rate usually does not exceed 100 packets/s (i.e., one packet per 100 frames for the case with frame duration of 0.1 ms) in MTC\cite{Khabazian2013tits380}, which means a sensor stays silent during 99\% of time. This indicates that, after a sensor has transmitted its packet, it will not generate new one and stay silent for a long period.} $\left|\mathcal{A}_{k}\right|$ is the cardinality of set $\mathcal{A}_{k}$.
	
	The inter-arrival time between packets at each sensor is equal to or higher than one frame, hence the UL queueing delay is always zero. The inter-arrival time between packets at BS could be shorter than one frame, hence the packets need to wait in the associated queue of the BS, and the DL queueing delay is non-zero. 
	
	To analyze queueing delay, effective bandwidth as a useful tool has been applied in the existing literature. The concept of effective bandwidth is the minimum constant packet service rate required to serve a random arrival process, which ensures queueing delay bound $D_{\mathrm{max}}^{\mathrm{q}}$ and queueing delay violation probability $\varepsilon^{\mathrm{q}}$. According to the result in \cite{She2018twc127}, the effective bandwidth of a Poisson process can be expressed as follows:
	\begin{equation}
		E_{k}^{\mathrm{B}}=\frac{T_\mathrm{f} \ln \left(1 / \varepsilon^{\mathrm{q}}\right)}{D_{\mathrm{max}}^{\mathrm{q}} \ln \left[\frac{T_\mathrm{f} \ln \left(1 / \varepsilon^{\mathrm{q}}\right)}{\lambda_{k} D_{\mathrm{max}}^{\mathrm{q}}}+1\right]}\ \left(\text{packets/frame}\right)\label{eq:effctive bandwidth},
	\end{equation}
	where $T_\mathrm{f}$ is the frame duration. It is widely believed that effective bandwidth is only applicable for the scenarios where the queueing delay bound is long\cite{Chen1995jsac1091}. However, the results in\cite{Choudhury1996tcomm203} show that, for Poisson process and the arrival processes which are more bursty than Poisson process, on condition of a short queueing delay bound, the approximated $\varepsilon^{\mathrm{q}}$ derived from effective bandwidth is an exact upper bound of queueing delay violation probability. This implies that effective bandwidth can be applied in URLLC. Further, \cite{She2018twc127} validates this implication with some typical arrival processes in URLLC, such as Poisson process, interrupted Poisson processes, and Switched Poisson process. 
	
	\subsection{QoS Requirements}
	\begin{figure}[!t]
		\centering
		\includegraphics[width=3.5in]{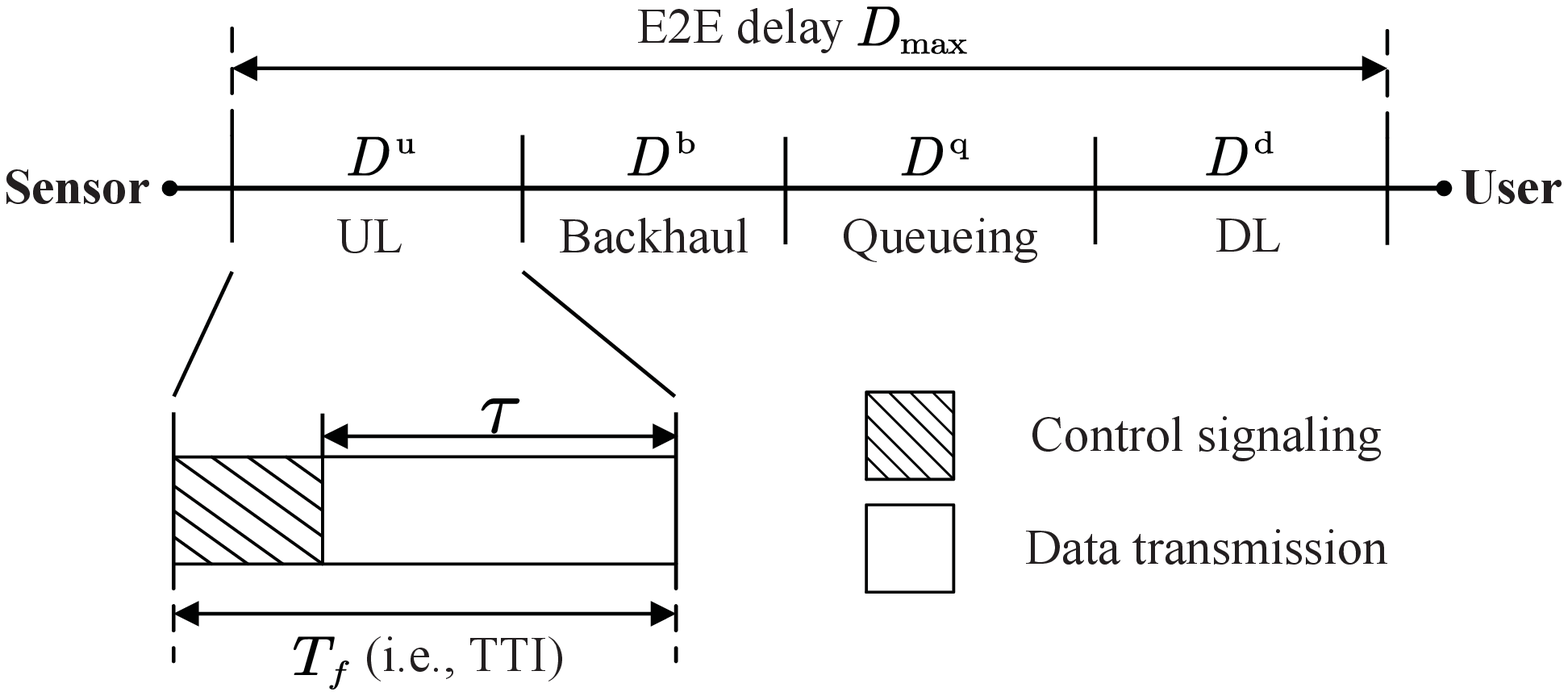}
		\caption{E2E latency and frame structure.}
		\label{fig:E2Edelay}
	\end{figure}
	The QoS requirements of URLLC is characterized by E2E latency $D_{\mathrm{max}}$ and overall packet loss probability $\varepsilon_{\mathrm{max}}$ that are imposed on each packet. As specified in \cite{TS22.261}, E2E latency refers to the time spent on delivering a packet from a source to a destination, measured at the air interface. For the considered local communications scenario, the over-the-air propagation delay can be ignored, and the latency for backhaul in fiber link is much shorter than 1 ms\cite{Zhang2016tcomm876}. Since the subchannel pipes are reserved for sensors, the accessing delay can also be ignored by contention-free random access with grant-free scheme. Therefore, as illustrated in Fig. \ref{fig:E2Edelay}, the latency excluding accessing delay in radio access network dominates E2E latency. The E2E latency is bounded by $D_{\mathrm{max}}$, and composed of UL and DL transmission delays, queueing delay at the buffer of BSs and backhaul delay. To reduce transmission delay, the short frame structure using mini-slot mode is considered\cite{Ashraf2015GC}. As shown in Fig. \ref{fig:E2Edelay}, each frame consists of two parts: control signaling and data transmission. In the control signaling, a part of overhead comes from CSI feedback for estimating full CSI.
	
	In UL and DL, each packet is finished within one frame. 
	Let $D^{\mathrm{u}}$, $D^{\mathrm{d}}$, $\varepsilon^{\mathrm{c},\mathrm{u}}$ and $\varepsilon^{\mathrm{c},\mathrm{d}}$ denote UL and DL transmission delays, UL and DL decoding error probabilities, respectively. If a packet is not transmitted error-free, then it will be lost. Let $D^{\mathrm{b}}$ and $D^{\mathrm{q}}$ denote the backhaul delay and queueing delay. The queueing delay for each packet should be bounded by $D_{\mathrm{max}}^{\mathrm{q}}\triangleq D_{\mathrm{max}}-D^{\mathrm{u}}-D^{\mathrm{d}}-D^{\mathrm{b}}$. If the duration that a packet spends on queueing exceeds $D_{\mathrm{max}}^{\mathrm{q}}$, then this packet is useless and has to be discarded. Therefore, the overall packet loss comes from decoding errors of UL and DL transmissions and the delay violation of queueing.
	
	In summary, the QoS requirements of URLLC can be satisfied under the following two constraints:
	\begin{equation}
		D^{\mathrm{u}}+D^{\mathrm{d}}+D^{\mathrm{q}}+D^{\mathrm{b}}\leq D_{\mathrm{max}},\label{eq:E2E latency}
	\end{equation}
	\begin{equation}
		1-(1-\varepsilon^{\mathrm{c},\mathrm{u}})(1-\varepsilon^{\mathrm{c},\mathrm{d}})(1-\varepsilon^{\mathrm{q}}) \leq \varepsilon_{\mathrm{max}},\label{eq:loss probability}
	\end{equation}
	where (\ref{eq:E2E latency}) and (\ref{eq:loss probability}) ensure latency and reliability, respectively. 
	
	\section{Packet Delivery Mechanism}\label{III}
	In this section, we focus on link layer design to cope with the chanllenge on ensuring target reliability of URLLC when channel stays in deep fading. We first summarize some current transmission schemes that try achieving the target reliability of URLLC. Then, we propose a packet delivery mechanism using frequency-hopping and proactive dropping, which makes URLLC more adaptive confronting of deep fading in MTC.
	\begin{figure}[!t]
		\centering
		\includegraphics[width=3.5in]{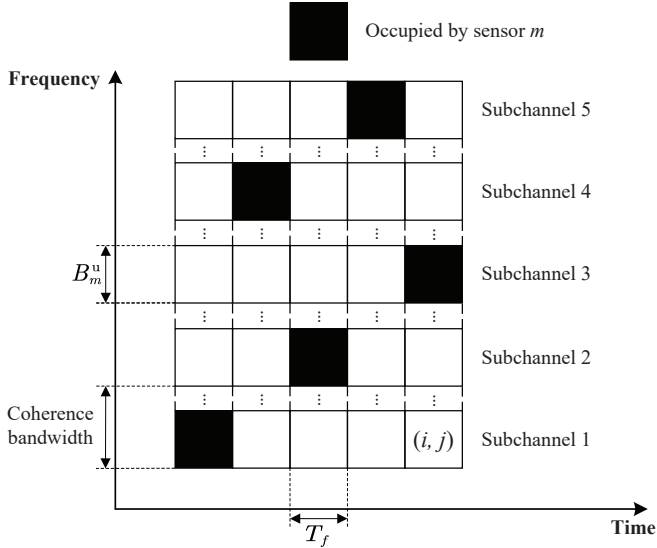}
		\caption{Exploiting frequency diversity by frequency-hopping.}
		\label{fig:Frequency-hopping}
	\end{figure}
	\subsection{Possible Transmission Schemes}
	To ensure the reliability, the received SNR must be equal to or higher than an SNR threshold.
	However, when channel is in deep fading, in order that the received SNR achieves the SNR threshold, the transmit power may need to be very high\cite{She2018twc127}.\footnote{The received SNR mainly depends on the change of small-scale channel gain since other channel parameters stay constant. Therefore, the transmit power gets unbounded when channel is in deep fading, i.e., $g\to 0$.} Unfortunately, subject to the constraint of maximum transmit power of a device or BS, so the target reliability of URLLC cannot be ensured. To this end, some transmission schemes have been employed in the literature or by the industry, but each has its own limitations.
	
	\begin{itemize}
		\item {\bf{HARQ:}} HARQ mechanism cannot be used to improve reliability in the event of a deep and lasting fade. This is because simply retransmitting a packet on the same radio resource in subsequent frames not only introduces additional latency, but also hardly improves the successful transmission probability when the channels stay in deep fading over multiple frames.
		
		\item {\bf{Proactive dropping:}} Proactive dropping mechanism proposed in \cite{She2018twc127} drops several packets in the queue under deep fading to control the overall reliability.\footnote{Another service policy is to drop all packets from the queue (queue clearing) to guarantee the overall reliability, but this will cause an outage of transmission.} However, now that one packet is transmitted in UL, the proactive dropping policy is equal to queue clearing and casues an outage of transmission. This makes the target reliability difficult to get satisified especially in UL.
		
		\item {\bf{Frequency diversity:}} Frequency diversity may not be scalable to the large number of nodes due to the limited bandwidth. Also, the assignment of subchannels introduces extra noise power and lowers the SNR, making the reliability harder to guarantee. 
		
		\item {\bf{Time diversity:}} Time diversity cannot be exploited to enhance reliability since channel remains constant within the coherence interval and varies independently among intervals, and hence utilizing time diversity exceeds the requirement of E2E latency.
		
		\item {\bf{Spatial diversity:}} Increasing the number of transmitting and/or receiving antennas can enhance the received SNR, and as such can also reduce the required transmit power\cite{Johansson2015ICCW},\cite{Tse2005}. But for the BS on high tower, larger antenna separation may be required. Hence, using spatial diversity with the moderate number of antennas can be an effctive way to improve reliability.
	\end{itemize}
	\begin{figure}[!t]
		\centering
		\includegraphics[width=3.5in]{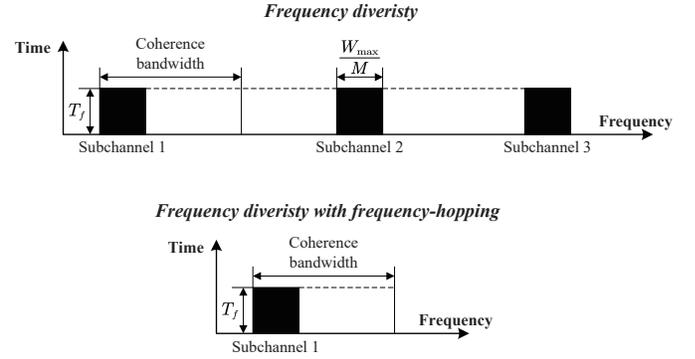}
		\caption{Bandwidth comparison.}
		\label{fig:bandwidth}
	\end{figure}
	\subsection{Proposed Delivery Mechanism}
	Considering the limitations of above-mentioned schemes, we propose a packet delivery mechanism that suits URLLC transmission in MTC better. To reduce the outage probability resulting from deep fading during URLLC transmission, we consider frequency-hopping for UL and proactive dropping for DL. Furthermore, since the available cellular spectrum is limited, both frequency-hopping and proactive dropping can help reduce the required bandwidth so that more devices can join in transmission in MTC.
	
	\subsubsection{Frequency-Hopping for UL}
	The packet can be transmitted over different subchannels in different transmission phases by introducing frequency-hopping. Specifically, when a sensor is accessed to the BS, a virtual subchannel pipe is reserved for it. With the pre-configured reservation scheme, there is no scheduling procedure before the transmission of each packet. The virtual subchannel pipe is a sequence of physical subchannels that will be occupied by a sensor in the next few frames\cite{Tse2005}, where the indices of the subchannels depend on the hopping pattern.\footnote{To avoid the interference, orthogonal virtual subchannel pipes should be reserved for different sensors in a hopping pattern. The design of hopping patterns between cells for collision avoidance should also be aware\cite{Tse2005}.} During the transmission, if the assigned subchannel in the current frame stays in deep fading, the sensor proactively quits transmitting in the current frame, and then transmits the packet in the subsequent frame over another subchannel according to the hopping pattern (if the assigned subchannel in the subsequent frame is not in deep fading)\cite{She2016GLOCOMW}.
	
	To obtain diversity gain, the separation between subchannels should be larger than the channel coherence bandwidth, such that the small-scale channel gains over subchannels are independent. Subject to E2E latency requirement of URLLC, the allowed maximum number of subchannels is denoted as $N_\mathrm{a}^\mathrm{max}$. An example of exploiting frequency diversity by frequency-hopping is illustrated in Fig. \ref{fig:Frequency-hopping}. The subchannels allocated to sensor $m$ is [(1,1), (2,4), (3,2), (4,5), (5,3)] with $N_\mathrm{a}^\mathrm{max}$ of 5, where the index of element represents the $j$-th subchannel that is allocated to sensor $m$ in the $i$-th frame, and the bandwdith of each subchannel is within the channel coherence bandwidth.
	
	Moreover, frequency-hopping contributes to bandwdith saving. In URLLC, several replicas of the packet with diversity technique are needed to improve the reliability, such as the proposed frequency diversity in \cite{She2018tcomm2266}. In contrast, frequency-hopping transmits the packet over the prespecified subchannels and frames based on hopping pattern. This can also improve the reliability of URLLC without the replicas, so the required bandwidth will be saved. An example of bandwidth comparsion between frequency-hopping and frequency diversity is illustrated in Fig. \ref{fig:bandwidth}, where $W_\mathrm{max}$ is the maximum available bandwidth. Assume that $W_\mathrm{max}$ is equally allocated to $M$ sensors, the bandwidth of frequency-hopping is exactly equal to $W_\mathrm{max}/M$. On the contrary, since the packet is repeatedly transmitted over multiple subchannels (e.g., 3 subchannels) concurrently, the required bandwidth of frequency diversity is triple that of $W_\mathrm{max}/M$. This implies that frequency-hopping helps reduce the required bandwidth and is expected to support more devices for URLLC.
	
	\subsubsection{Proactive Dropping for DL}
	The core idea of proactive packet dropping is to discard several packets from the queue to lower the required SNR, such that decoding error probability can be satisfied with finite transmit power. Although some of packets can not yet be served with $D_{\mathrm{max}}^{\mathrm{q}}$ and $\varepsilon^{\mathrm{q}}$, by further controlling the proactive dropping probability with resource allocation, the overall packet loss probability can be ensured no matter how these packets are lost. In this paper we adopt the service policy proposed in \cite{She2018twc127}, where only a part of the $E_{k}^{\mathrm{B}}$ packets are proactively discarded when the channel is in deep fading. Compared to the original policy of queue clearing that is used for ensuring reliability, the proactive dropping avoids the outage of transmission. Besides, under the same target reliability of URLLC, the reuqired bandwidth of proactive dropping can be further saved since several packets has been discarded from the queue, so that more devices can be supported.
	
	\section{Problem Formulation}\label{IV}
	In this section, we formulate a resource allocation problem that minimizes the average total power consumption under the QoS requirements of URLLC and the constraint of maximum transmit power. We first introduce how the kinds of packet loss components of UL and DL are ensured by the proposed delivery mechanism. Then we derive the expression of EE of URLLC and represent the equivalency between maximizing EE and minimizing power consumption. Frequently used notations throughout the paper are summarized in Table \ref{tab:notations}.
	\begin{table*}[!t]
		\caption{Summary of notations}\label{tab:notations}
		\centering
		\renewcommand\arraystretch{1.2}{
			\setlength{\tabcolsep}{2mm}{
				\begin{tabular}{|p{1.8cm}<{\centering}|p{6.4cm}<{\raggedright}||p{1.8cm}<{\centering}|p{6.4cm}<{\raggedright}|}
					\hline
					$M$& Number of sensors& $K$& Number of users\\
					\hline
					$\mathcal{S}$ & Set of sensors that are associated to the BS& $\mathcal{U}$& Set of users that are associated to the BS\\
					\hline
					$\kappa$& Active probability of each sensor during one frame& $\xi_{k}$& Nonempty probability of the queue to the $k$-th user\\
					\hline
					$\mathcal{A}_{k}$& Set of active sensors that transmit pakctes to the $k$-th user & $\lambda_{k}$& Average packet arrival rate of Poisson process to the $k$-th user\\
					\hline
					$N_\mathrm{t}$& Number of antennas& $T_f$& Duration of one frame\\
					\hline
					$\tau$& Duration of data transmission & $L$& Pakcet size\\
					\hline
					$\phi$& SNR loss&  $N_0$& Single-sided noise spectral density\\
					\hline
					$W_\mathrm{max}$& Maximum available bandwidth & $W_\mathrm{c}$& Channel coherence bandwidth\\
					\hline
					$N_\mathrm{a}^\mathrm{max}$& Allowed maximum number of subchannels&  $N_\mathrm{a}$& Number of assigned subchannels\\
					\hline
					$i$& Indice of the assigned subchannel&  $N_m^{\mathrm{u}}$& Number of waiting frames before delivering the error-free packet from the $m$-th sensor to the BS\\
					\hline
					$D_\mathrm{max}$& E2E latency requirement& $\varepsilon_\mathrm{max}$& Overall packet loss probability\\
					\hline
					$D^\mathrm{u}_\mathrm{max}$& UL transmission delay bound& $D^\mathrm{q}_\mathrm{max}$& Queueing delay bound\\
					\hline
					$\varepsilon^\mathrm{q}$& Requirement of queueing delay violation probability& $E_{k}^{\mathrm{B}}\left(N_{\mathrm{a}}\right)$& Effective bandwidth of Poisson process to the $k$-th user\\
					\hline
					$\varepsilon^\mathrm{c,u}$, $\varepsilon^\mathrm{c,d}$& Requirements of decoding error probability of UL and DL& $\varepsilon^\mathrm{p,u}$, $\varepsilon^\mathrm{p,d}$& Requirement of packet dropping probability of UL and DL\\
					\hline
					$\mu_m^\mathrm{u}$, $\mu_k^\mathrm{d}$& Large-scale channel gain of the $m$-th sensor and the $k$-th user& $B_m^\mathrm{u}$, $B_k^\mathrm{d}$& Allocated bandwidth of the $m$-th sensor and the $k$-th user\\
					\hline
					$g_{m,i}^{\mathrm{u}}$, $g_{k}^{\mathrm{d}}$&  Small-scale channel gain of the $m$-th sensor over the $i$-th subchannel and the $k$-th user& $P_{m,i}^{\mathrm{u}}$, $P_{k}^{\mathrm{d}}$& Allocated transmit power of the $m$-th sensor over the $i$-th subchannel and the $k$-th user\\
					\hline
					$\gamma_m^{\mathrm{th},{\mathrm{u}}}$, $\gamma_k^{\mathrm{th},{\mathrm{d}}}$& SNR threshold of the $m$-th sensor and the $k$-th user& $P_m^{\mathrm{th},{\mathrm{u}}}$, $P_k^{\mathrm{th},{\mathrm{d}}}$& Transmit power threshold of the $m$-th sensor and the $k$-th user\\
					\hline
					$g_m^{\mathrm{th},{\mathrm{u}}}$, $g_k^{\mathrm{th},{\mathrm{d}}}$& Small-scale channel gain threshold of the $m$-th sensor and the $k$-th user& $P_{\mathrm{max}}^{\mathrm{u}}$, $P_{\mathrm{max}}^{\mathrm{d}}$& Maximum transmit power of a sensor and a BS\\
					\hline
					$\rho^\mathrm{u}$, $\rho^\mathrm{d}$& Power amplifier efficiency of each sensor and BS&  $P_{\mathrm {tot}}^{\mathrm {UB}}$& Upper bound of average total power consumption\\
					\hline
					$W_m^\mathrm{th,u}$, $W_k^\mathrm{th,d}$& Stationary point of bandwidth of the $m$-th sensor and the $k$-th user&  $N_{\mathrm{t}}^{\mathrm{min}}$& Required minimum number of antennas that satisfies the overall reliability with the proposed scheme\\
					\hline	
		\end{tabular}}}
	\end{table*}
	\subsection{Ensuring QoS Constraints}
	
	\subsubsection{UL Packet Loss Probability}
	Denote $N_{\mathrm{a}}$ as the number of assigned subchannels in a virtual subchannel pipe, then the achievable rate from the $m$-th sensor to the BS over the $i$-th subchannel ($1\leq i\leq N_{\mathrm{a}}, i\in \mathbb{N}$) is given by
	\begin{equation}
		\begin{aligned}
			R_{m,i}^{\mathrm{u}}= \frac{\tau B_m^{\mathrm{u}}}{L\ln2}\left[\ln\left(1+\frac{\mu_m^{\mathrm{u}} g_{m,i}^{\mathrm{u}} P_{m,i}^{\mathrm{u}}}{\phi N_0 B_m^{\mathrm{u}}}\right)-\frac{f_Q^{-1}\left(\varepsilon_{m,i}^{\mathrm{c},\mathrm{u}}\right)}{\sqrt{\tau B_m^{\mathrm{u}}}}\right]\\
			,m\in\mathcal{S}\ \left(\text{packets/frame}\right),\label{eq:ULrate}
		\end{aligned}
	\end{equation}
	where $\mathcal{S}$ denotes the set of sensors that are associated to the BS, $B_m^{\mathrm{u}}$ and $\mu_m^{\mathrm{u}}$ are the allocated bandwidth and large-scale channel gain of the $m$-th sensor, $P_{m,i}^{\mathrm{u}}$, $g_{m,i}^{\mathrm{u}}$ and $\varepsilon_{m,i}^{\mathrm{c},\mathrm{u}}$ are the transmit power, small-scale channel gain and decoding error probability of the $m$-th sensor over the $i$-th subchannel, respectively. Since one packet is uploaded within one frame, the required SNR threshold over each subchannel can be obatined by substituting (\ref{eq:ULrate}) into $R_{m,i}^{\mathrm{u}}=1$:
	\begin{equation}
		\gamma_{m}^{\mathrm{th},\mathrm{u}}\triangleq \exp\left[\frac{L \ln2}{\tau B_m^{\mathrm{u}}}+\frac{f_Q^{-1}\left(\varepsilon^{\mathrm{c},\mathrm{u}}\right)}{\sqrt{\tau B_m^{\mathrm{u}}}}\right]-1,\label{eq:UL_SNR}
	\end{equation}
	where $\varepsilon^{\mathrm{c},\mathrm{u}}$ is the UL decoding error probability reuqirement.
	
	Frequency-hopping transmits the packet when the channel does not stay in deep fading, which indicates that the received SNR $\gamma_{m,i}^{\mathrm{u}}$ is equal to or higher than $\gamma_{m}^{\mathrm{th},\mathrm{u}}$. However, sometimes the subchannels in the next few frames are all good enough for transmission, such that the total power gets accumulated. To save energy, we only allow each sensor to transmit the packet once only at the first time that the channel is not in deep fading, and it will then stay silent even if another well-conditioned channel appears.
	
	Let $P_{\mathrm{max}}^{\mathrm{u}}$ and $P_m^{\mathrm{u}}$ denote the maximum transmit power of each sensor and transmit power of the $m$-th sensor, respectively. To ensure $P_m^{\mathrm{u}}\leq P_{\mathrm{max}}^{\mathrm{u}}$, the transmit power threshold $P_m^{\mathrm{th},{\mathrm{u}}}$ is introduced to the $m$-th sensor where $P_m^{\mathrm{th},{\mathrm{u}}}\leq P_{\mathrm{max}}^{\mathrm{u}}$\cite{She2018twc127}, and control $P_m^{\mathrm{u}}\leq P_m^{\mathrm{th},\mathrm{u}}$. Let $N_m^{\mathrm{u}}$ denote the number of waiting frames that ensures the packet to be transmitted successfully. As a result, $D^{\mathrm{u}}=\left(N_m^{\mathrm{u}}+1\right)T_f$ and should be bounded by $D_{\mathrm{max}}^{\mathrm{u}}\triangleq N_{\mathrm{a}}T_f$. Depending on the hopping pattern and channel condition, $N_m^{\mathrm{u}}$ is a random variable and then set $N_m^{\mathrm{u}}=j$ where $j\in \left[0, N_{\mathrm{a}}-1\right], j\in \mathbb{N}$. 
	
	When UL transmission delay exceeds $D_{\mathrm{max}}^{\mathrm{u}}$ (i.e., $N_m^{\mathrm{u}}\geq N_{\mathrm{a}}$), then a packet becomes invalid since it violates the latency requirement, and thus no power is transmitted from the sensor. In the case of $N_m^{\mathrm{u}}<N_{\mathrm{a}}$, UL decoding error probability requirement can be satisfied, and then \emph{channel inversion} is applied to achieve the required received SNR in (\ref{eq:UL_SNR}). Then, when the $m$-th sensor has a packet to transmit, the power control policy can be expressed as follows:
	\begin{equation}
		P_{m}^{\mathrm{u}}=
		\begin{cases}
			0, &\text{if } N_m^{\mathrm{u}}\geq N_{\mathrm{a}},\\
			\frac{\phi N_{0} B_{m}^{\mathrm{u}} \gamma_{m}^{\mathrm{th},\mathrm{u}}}{\mu_{m}^{\mathrm{u}} g_{m,j+1}^{\mathrm{u}}}, &\text{if } N_m^{\mathrm{u}}= j.
		\end{cases}\label{eq:ULpower1}\\
	\end{equation}
	
	In fact, whether the sensor proactively quits transmitting over the current assigned subchannel depends on small-scale channel gain, hence (\ref{eq:ULpower1}) can be equivalently expressed as
	\begin{equation}
		P_{m}^{\mathrm{u}}=
		\begin{cases}
			0, & \text{if } \bigcap_{i=1}^{N_a}\left(g_{m,i}^{\mathrm{u}}<g_{m}^{\mathrm{th},\mathrm{u}}\right),\\
			\frac{\phi N_{0} B_{m}^{\mathrm{u}} \gamma_{m}^{\mathrm{th},\mathrm{u}}}{\mu_{m}^{\mathrm{u}} g_{m,j+1}^{\mathrm{u}}}, & \text{if }\bigcap_{i=1}^{j}\left(g_{m,i}^{\mathrm{u}}<g_{m}^{\mathrm{th},\mathrm{u}}\right),g_{m,j+1}^{\mathrm{u}}\geq g_{m}^{\mathrm{th},\mathrm{u}},
		\end{cases}\label{eq:ULpower2}
	\end{equation}
	where the threshold $g_m^{\mathrm{th},\mathrm{u}}$ is defined as
	\begin{equation}
		g_m^{\mathrm{th},\mathrm{u}}\triangleq \frac{\phi N_0 B_m^{\mathrm{u}}\gamma_{m}^{\mathrm{th},\mathrm{u}}}{\mu_m^{\mathrm{u}}P_{m}^{\mathrm{th},\mathrm{u}}}.
		\label{eq:ULgainthreshold}
	\end{equation}
	
	As stated in the policy, a packet will be discared when $N_m^{\mathrm{u}}\geq N_{\mathrm{a}}$. Then, the UL packet dropping probability of the $m$-th sensor is given by
	\begin{equation}
		\begin{aligned}
			B_{N_\mathrm{t},N_\mathrm{a}}^{\mathrm{u}}\left(g_{m}^{\mathrm{th}, \mathrm{u}}\right)&\triangleq\left[\int_{0}^{g_{m}^{\mathrm{th}, \mathrm{u}}}f_{N_\mathrm{t}}(g) \mathrm{d} g\right]^{N_{\mathrm{a}}}\\
			&=\left[1-e^{-g_{m}^{\mathrm{th}, \mathrm{u}}} \sum_{n=0}^{N_\mathrm{t}-1} \frac{\left(g_{m}^{\mathrm{th}, \mathrm{u}}\right)^{n}}{n!}\right]^{N_{\mathrm{a}}}.
		\end{aligned}\label{eq:ULdroppingFORMULA}
	\end{equation}
	In UL transmission, the packet dropping probability is equal to the outage probability. Compared to the outage probability in \cite{She2019twc402}, (\ref{eq:ULdroppingFORMULA}) decreases exponentially by introducing more subchannels. To guarantee UL packet dropping probability, the following constraint should be satisfied:
	\begin{equation}
		B_{N_\mathrm{t},N_\mathrm{a}}^{\mathrm{u}}\left(\frac{\phi N_0 B_m^{\mathrm{u}}\gamma_m^{\mathrm{th},\mathrm{u}}}{\mu_m^{\mathrm{u}}P_{m}^{\mathrm{th},\mathrm{u}}}\right)=\varepsilon^{\mathrm{p}, \mathrm{u}},\label{eq:ULdropping}
	\end{equation}
	where $\varepsilon^{\mathrm{p},\mathrm{u}}$ is the UL packet dropping probability reuqirement.
	
	\subsubsection{DL Packet Loss Probability}
	The achievable rate from the BS to the $k$-th user is given by
	\begin{equation}
		\begin{aligned}
			R_k^{\mathrm{d}}= \frac{\tau B_k^{\mathrm{d}}}{L\ln2}\left[\ln\left(1+\frac{\mu_k^{\mathrm{d}} g_k^{\mathrm{d}} P_{k}^{\mathrm{d}}}{\phi N_0 B_k^{\mathrm{d}}}\right)-\frac{f_Q^{-1}\left(\varepsilon_{k}^{\mathrm{c},\mathrm{d}}\right)}{\sqrt{\tau B_k^{\mathrm{d}}}}\right]\\
			,k\in\mathcal{U}\ \left(\text{packets/frame}\right),
		\end{aligned}\label{eq:DLrate}
	\end{equation}
	where $\mathcal{U}$ denotes the set of users associated to the BS, $B_k^{\mathrm{d}}$ and $P_{k}^{\mathrm{d}}$ are the allocated bandwidth and transmit power of the $k$-th user, $\mu_k^{\mathrm{d}}$ and $ g_k^{\mathrm{d}}$ are the large-scale and small-scale channel gain of the $k$-th user. Since UL transmission is bounded by $D_{\mathrm{max}}^{\mathrm{u}}$, DL transmission delay and backhaul delay respectively take one frame, then the queueing delay bound can be obatained as follows:
	\begin{equation}
		\begin{aligned}
			D_{\mathrm{max}}^{\mathrm{q}}&= D_{\mathrm{max}}-D_{\mathrm{max}}^{\mathrm{u}}-D^{\mathrm{d}}-D^{\mathrm{b}}\\
			&=D_{\mathrm{max}}-\left(N_{\mathrm{a}}+2\right)T_f.\label{eq:delay bound}
		\end{aligned}
	\end{equation}
	Substituting (\ref{eq:delay bound}) and $\varepsilon^{\mathrm{q}}$ into (\ref{eq:effctive bandwidth}), the effective bandwidth is given by
	\begin{equation}
		\begin{aligned}
			&E_{k}^{\mathrm{B}}\left(N_{\mathrm{a}}\right)\\
			=&\frac{T_\mathrm{f} \ln \left(1 / \varepsilon^{\mathrm{q}}\right)}{\left[D_{\mathrm{max}}-\left(N_{\mathrm{a}}+2\right)T_f\right] \ln \left\{\frac{T_\mathrm{f} \ln \left(1 / \varepsilon^{\mathrm{q}}\right)}{\lambda_{k} \left[D_{\mathrm{max}}-\left(N_{\mathrm{a}}+2\right)T_f\right]}+1\right\}}.
		\end{aligned}
	\end{equation} 
	
	Since $D_{\mathrm{max}}^{\mathrm{q}}$ is shorter than the channel coherence time, the service rate is constant within the latency requirement with given resources\cite{She2018twc127}. To satisfy the requirements of $(D_{\mathrm{max}}^{\mathrm{q}}, \varepsilon^{\mathrm{q}})$ and $\varepsilon^{\mathrm{c},\mathrm{d}}$ that are imposed on each packet, the constant packet service rate should be no less than the effective bandwidth. Therefore, the required received SNR can be obtained by substituting $E_{k}^{\mathrm{B}}$ into $R_k^{\mathrm{d}}= E_{k}^{\mathrm{B}}\left(N_{\mathrm{a}}\right)$:
	\begin{equation}
		\gamma_{k}^{\mathrm{th},\mathrm{d}}\triangleq \exp\left[\frac{E_{k}^{\mathrm{B}}\left(N_{\mathrm{a}}\right) L\ln2}{\tau B_k^{\mathrm{d}}}+\frac{f_Q^{-1}\left(\varepsilon^{\mathrm{c},\mathrm{d}}\right)}{\sqrt{\tau B_k^{\mathrm{d}}}}\right]-1.\label{eq:DL_SNR}
	\end{equation}
	
	Let $P_{\mathrm{max}}^{\mathrm{d}}$ and $P_k^{\mathrm{d}}$ denote the maximum transmit power of BS and allocated transmit power to the $k$-th user, respectively. To ensure $\sum_{k\in\mathcal{U}}P_k^{\mathrm{d}}\leq P_{\mathrm{max}}^{\mathrm{d}}$, the transmit power threshold $P_k^{\mathrm{th},\mathrm{d}}$ is introduced to the $k$-th user where $\sum_{k\in\mathcal{U}}P_k^{\mathrm{th},\mathrm{d}}\leq P_{\mathrm{max}}^{\mathrm{d}}$\cite{She2018twc127}, and control $P_k^{\mathrm{d}}\leq P_k^{\mathrm{th},\mathrm{d}}$.
	
\begin{figure}[!t]
		\centering
		\includegraphics[width=3.6in]{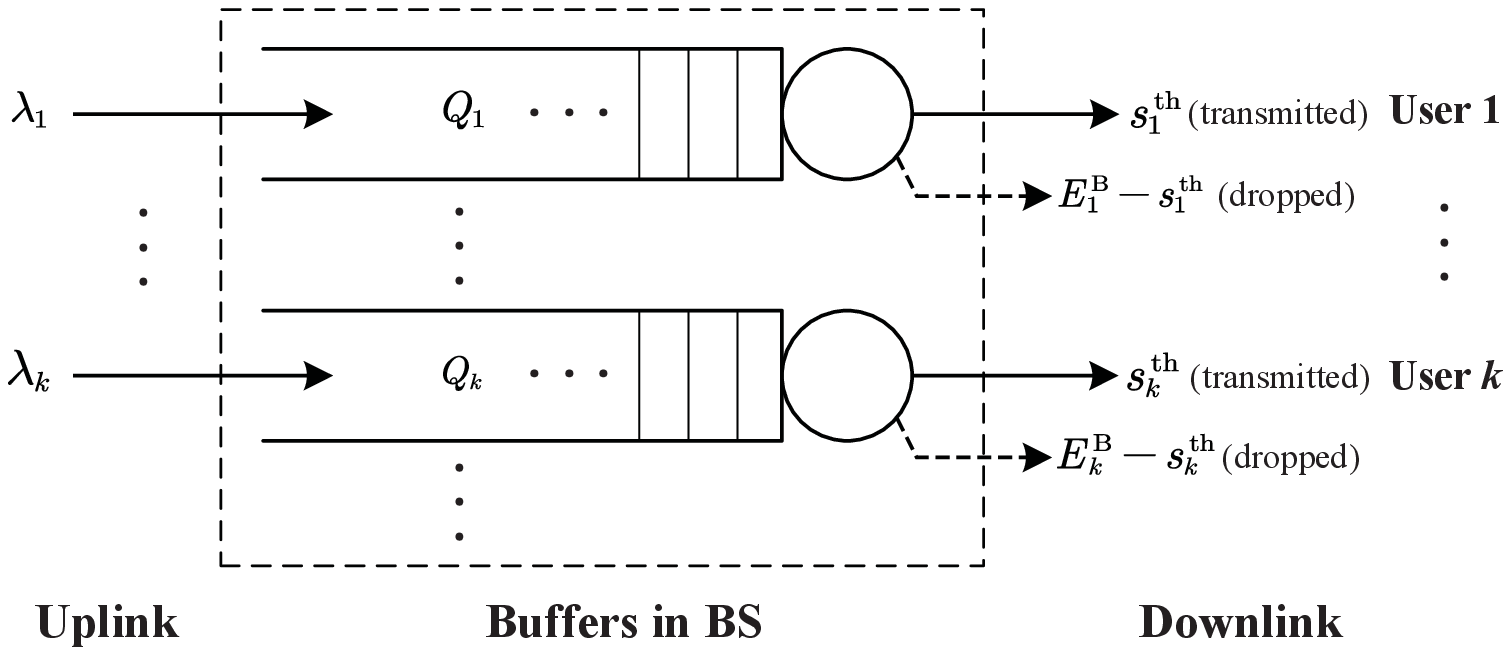}
		\caption{Proactive packet dropping scheme.}
		\label{fig:proactivedropping}
	\end{figure}

	When the channel is in deep fading, the received SNR can be lower than $\gamma_{k}^{\mathrm{th},\mathrm{d}}$ with finite transmit power. In this case of $R_k^{\mathrm{d}}<E_{k}^{\mathrm{B}}\left(N_{\mathrm{a}}\right)$, not all packets in the queue can be served under the requirements of $(D_{\mathrm{max}}^{\mathrm{q}}, \varepsilon^{\mathrm{q}})$ and $\varepsilon^{\mathrm{c},\mathrm{d}}$ in the current frame. However, some of packets can still be served under $\varepsilon^{\mathrm{c},\mathrm{d}}$ after reducing the number of transmitted packets. To guarantee the target reliability of URLLC with finite transmit power, the scheme in \cite{She2018twc127} controls the overall reliability by proactively dropping several packets in the queue under deep fading. As illustrated in Fig. \ref{fig:proactivedropping}, some packets are transmitted at rate $s_k^{\mathrm{th}}$, which is obtained by substituting $P_k^{\mathrm{th},\mathrm{d}}$ into (\ref{eq:DLrate}). The rest of the packets are dropped at rate $b_{k}=\min \left\{E_{k}^{\mathrm{B}}\left(N_{\mathrm{a}}\right)-s_k^{\mathrm{th}},0\right\}$ when the queue length of $k$-th user $Q_k$ exceeds $E_{k}^{\mathrm{B}}\left(N_{\mathrm{a}}\right)$\cite{She2018twc127}. If there is no packet for the $k$-th user in the queue at the BS, then no power will be allocated to the user. Otherwise, (\ref{eq:DL_SNR}) should be satisfied after proactive dropping.
	
	Then, when the queue for the $k$-th user is nonempty, the power control policy can be expressed as follows:
	\begin{align}
		&P_{k}^{\mathrm{d}}=
		\begin{cases}
			P_{k}^{\mathrm{th},\mathrm{d}}, &\text{if }g_{k}^{\mathrm{d}}<g_{k}^{\mathrm{th},\mathrm{d}},\\
			\frac{\phi N_{0} B_{k}^{\mathrm{d}} \gamma_{k}^{\mathrm{th},\mathrm{d}}}{\mu_{k}^{\mathrm{d}} g_{k}^{\mathrm{d}}}, &\text{if }g_{k}^{\mathrm{d}}\geq g_{k}^{\mathrm{th},\mathrm{d}},
		\end{cases}\label{eq:DLpower}
	\end{align}
	where the threshold $g_k^{\mathrm{th},\mathrm{d}}$ is defined as
	\begin{equation}
		g_k^{\mathrm{th},\mathrm{d}}\triangleq \frac{\phi N_0 B_k^{\mathrm{d}}\gamma_{k}^{\mathrm{th},\mathrm{d}}}{\mu_k^{\mathrm{d}}P_{k}^{\mathrm{th},\mathrm{d}}}.
		\label{eq:DLgainthreshold}
	\end{equation}
	
	According to the policy, the DL proactive packet dropping probability of the $k$-th user is bounded by\cite{She2019twc402}
	\begin{equation}
		\begin{aligned}
			B_{N_\mathrm{t},N_\mathrm{a}}^{\mathrm{d}}\left(g_{k}^{\mathrm{th}, \mathrm{d}}\right) \triangleq & \int_{0}^{g_{k}^{\mathrm{th}, \mathrm{d}}}\left(1-\frac{g}{g_{k}^{\mathrm{th}, \mathrm{d}}}\right) f_{N_\mathrm{t}}(g) \mathrm{d} g \\
			=&\left(1-\frac{N_\mathrm{t}}{g_{k}^{\mathrm{th}, \mathrm{d}}}\right) e^{-g_{k}^{\mathrm{th}, \mathrm{d}}} \sum_{n=0}^{N_\mathrm{t}-1} \frac{\left(g_{k}^{\mathrm{th}, \mathrm{d}}\right)^{n}}{n !} \\
			&+e^{-g_{k}^{\mathrm{th}, \mathrm{d}}} \frac{\left(g_{k}^{\mathrm{th}, \mathrm{d}}\right)^{N_\mathrm{t}-1}}{\left(N_\mathrm{t}-1\right) !}.
		\end{aligned}\label{eq:DLdroppingFORMULA}
	\end{equation}
	To guarantee DL proactive packet dropping probability, the following constraint should be satisfied:
	\begin{equation}
		B_{N_\mathrm{t},N_\mathrm{a}}^{\mathrm{d}}\left(\frac{\phi N_0 B_k^{\mathrm{u}}\gamma_{k}^{\mathrm{th},\mathrm{d}}}{\mu_k^{\mathrm{d}}P_{k}^{\mathrm{th},\mathrm{d}}}\right)=\varepsilon^{\mathrm{p}, \mathrm{d}},\label{eq:DLdropping}
	\end{equation}
	where $\varepsilon^{\mathrm{p},\mathrm{d}}$ is the DL packet dropping probability reuqirement.
	
	\subsubsection{Overall Packet Loss Probability}
	To guarantee QoS metrics of URLLC with finite transmit power under deep fading, we utilize frequency-hopping for UL and proactive dropping for DL. As detailed in this section, the transmission of both being error-free and not violating delay bound is required for UL and DL. Then, the overall packet loss probability that is imposed on each packet can be re-expressed as follows:
	\begin{equation}
		\begin{aligned}
			1-&(1-\varepsilon^{\mathrm{c},\mathrm{u}})(1-\varepsilon^{\mathrm{p},\mathrm{u}})(1-\varepsilon^{\mathrm{c},\mathrm{d}})(1-\varepsilon^{\mathrm{p},\mathrm{d}})(1-\varepsilon^{\mathrm{q}})\\
			&\approx\varepsilon^{\mathrm{c},\mathrm{u}}+\varepsilon^{\mathrm{p},\mathrm{u}}+\varepsilon^{\mathrm{c},\mathrm{d}}+\varepsilon^{\mathrm{p},\mathrm{d}}+\varepsilon^{\mathrm{q}} \leq \varepsilon_{\mathrm{max}},\label{eq:packet loss}
		\end{aligned}
	\end{equation}
	where the approximation is accurate since $\varepsilon^{\mathrm{c},\mathrm{u}}$, $\varepsilon^{\mathrm{p},\mathrm{u}}$, $\varepsilon^{\mathrm{c},\mathrm{d}}$, $\varepsilon^{\mathrm{p},\mathrm{d}}$ and $\varepsilon^{\mathrm{q}}$ are extremely small\cite{She2018twc127,She2018tcomm2266,She2019twc402}.
	
	\subsection{Objective Function}
	The average UL transmit power of the $m$-th sensor can be obtained as follows:
	\begin{equation}
		\begin{aligned}
			&\mathbb{E}\left\{P_{m}^{\mathrm{u}}\right\}\\
			=&\kappa\sum_{j=0}^{N_{\mathrm{a}}-1}\left[\int_{0}^{g_{m}^{\mathrm{th},\mathrm{u}}}f_{N_\mathrm{t}}(g)\mathrm{d}g\right]^j\int_{g_{m}^{\mathrm{th},\mathrm{u}}}^{\infty}\frac{\phi N_{0} B_{m}^{\mathrm{u}} \gamma_{m}^{\mathrm{th},\mathrm{u}}}{\mu_{m}^{\mathrm{u}} g_{m,j+1}^{\mathrm{u}}}f_{N_\mathrm{t}}(g)\mathrm{d}g\\
			=&\sum_{j=0}^{N_{\mathrm{a}}-1}\left[B_{N_\mathrm{t},N_\mathrm{a}}^{\mathrm{u}}\left(g_{m}^{\mathrm{th}, \mathrm{u}}\right)\right]^{\frac{j}{N_{\mathrm{a}}}}\frac{\kappa\phi N_{0} B_{m}^{\mathrm{u}} \gamma_{m}^{\mathrm{th},\mathrm{u}}}{\mu_{m}^{\mathrm{u}} \left(N_\mathrm{t}-1\right)}\int_{0}^{g_{m}^{\mathrm{th}, \mathrm{u}}} f_{N_\mathrm{t}-1}(g) \mathrm{d} g\\
			=&\sum_{j=0}^{N_{\mathrm{a}}-1}\left[B_{N_\mathrm{t},N_\mathrm{a}}^{\mathrm{u}}\left(g_{m}^{\mathrm{th}, \mathrm{u}}\right)\right]^{\frac{j}{N_{\mathrm{a}}}}\frac{\kappa\phi N_{0} B_{m}^{\mathrm{u}} \gamma_{m}^{\mathrm{th},\mathrm{u}}}{\mu_{m}^{\mathrm{u}} \left(N_\mathrm{t}-1\right)}\\
			&\qquad\  \cdot \left[1-\left(B_{N_\mathrm{t}-1,N_\mathrm{a}}^{\mathrm{u}}\left(g_{m}^{\mathrm{th}, \mathrm{u}}\right)\right)^{\frac{1}{N_{\mathrm{a}}}}\right]\\
			\leq&\sum_{j=0}^{N_{\mathrm{a}}-1}\left[B_{N_\mathrm{t},N_\mathrm{a}}^{\mathrm{u}}\left(g_{m}^{\mathrm{th}, \mathrm{u}}\right)\right]^{\frac{j}{N_{\mathrm{a}}}}\frac{\kappa\phi N_{0} B_{m}^{\mathrm{u}} \gamma_{m}^{\mathrm{th},\mathrm{u}}}{\mu_{m}^{\mathrm{u}} \left(N_\mathrm{t}-1\right)}.
		\end{aligned}\label{eq:ULpower_av}
	\end{equation}
	The upper bound in (\ref{eq:ULpower_av}) is very tight since $\varepsilon^{\mathrm{p}, \mathrm{u}}$ is extremely small in order to meet the ultra-high reliability requirement.
	
	The average DL transmit power allocated to the $k$-th user can be obtained as
	\begin{equation}
		\begin{aligned}
			&\mathbb{E}\left\{P_{k}^{\mathrm{d}}\right\}\\
			=&\xi_{k}\left[\int_{0}^{g_{k}^{\mathrm{th}, \mathrm{d}}} P_{k}^{\mathrm{th}, \mathrm{d}} f_{N_\mathrm{t}}(g) \mathrm{d} g+\int_{g_{k}^{\mathrm{th}, \mathrm{d}}}^{\infty} \frac{\phi N_{0} B_{k}^{\mathrm{d}} \gamma_{k}^{\mathrm{th},\mathrm{d}}}{\mu_{k}^{\mathrm{d}} g_{k}^{\mathrm{d}}} f_{N_\mathrm{t}}(g) \mathrm{d} g\right]\\
			=&\frac{\xi_{k} \phi N_{0} B_{k}^{\mathrm{d}} \gamma_{k}^{\mathrm{th},\mathrm{d}}}{\mu_{k}^{\mathrm{d}}\left(N_\mathrm{t}-1\right)}\left[\int_{0}^{g_{k}^{\mathrm{th}, \mathrm{d}}} \frac{g_{k}^{\mathrm{d}}}{g_{k}^{\mathrm{th}, \mathrm{d}}}f_{N_\mathrm{t}-1}(g) \mathrm{d} g\right]\\
			&+\frac{\xi_{k} \phi N_{0} B_{k}^{\mathrm{d}} \gamma_{k}^{\mathrm{th},\mathrm{d}}}{\mu_{k}^{\mathrm{d}}\left(N_\mathrm{t}-1\right)}\left[1- \int_{0}^{g_{k}^{\mathrm{th}, \mathrm{d}}} f_{N_\mathrm{t}-1}(g) \mathrm{d} g\right]\\
			=&\frac{\xi_{k} \phi N_{0} B_{k}^{\mathrm{d}} \gamma_{k}^{\mathrm{th},\mathrm{d}}}{\mu_{k}^{\mathrm{d}}\left(N_\mathrm{t}-1\right)}\left[1-B_{N_\mathrm{t},N_\mathrm{a}}^{\mathrm{d}}\left(g_{k}^{\mathrm{th}, \mathrm{d}}\right)\right]\\
			\leq&\frac{\xi_{k} \phi N_{0} B_{k}^{\mathrm{d}} \gamma_{k}^{\mathrm{th},\mathrm{d}}}{\mu_{k}^{\mathrm{d}}\left(N_\mathrm{t}-1\right)},
		\end{aligned}\label{eq:DLpower_av}
	\end{equation}
	where $\xi_{k}=\lambda_{k}/E_{k}^{\mathrm{B}}\left(N_{\mathrm{a}}\right)$ is the probability that the queue at BS for the $k$-th user is nonempty. The upper bound in (\ref{eq:DLpower_av}) is also very tight since $\varepsilon^{\mathrm{p}, \mathrm{d}}$ is extremely small in order to meet the ultra-high reliability requirement.
	
	The average total power consumed by the $m$-th sensor can be modelled as follows\cite{Arnold2010},\cite{Debaillie2015}:
	\begin{equation}
		\mathbb{E}\left\{P_{\mathrm {tot},m}^{\mathrm{u}}\right\}=\frac{1}{\rho^\mathrm{u}}\mathbb{E}\left\{P_{m}^{\mathrm{u}}\right\}+P^{\mathrm{c,u}},
	\end{equation}
	where $\rho^\mathrm{u}$ is the power amplifier (PA) efficiency of each sensor, $\mathbb{E}\left\{P_{m}^{\mathrm{u}}\right\}$ the average transmit power of the $m$-th sensor, and $P^{\mathrm{c},\mathrm{u}}$ the circuit power consumed at each sensor.
	
	The average total power consumed by the BS can be modelled as follows\cite{Arnold2010},\cite{Debaillie2015}:
	\begin{equation}
		\mathbb{E}\left\{P_{\mathrm {tot}}^{\mathrm{d}}\right\}=\frac{1}{\rho^\mathrm{d}}\sum_{k \in \mathcal{U}}\mathbb{E}\left\{P_{k}^{\mathrm{d}}\right\}+N_\mathrm{t}P^{\mathrm{c,nt}}+\frac{P^{\mathrm{c,na}}}{N_\mathrm{a}},
	\end{equation}
	where $\rho^\mathrm{d}$ is the PA efficiency of BS, $\mathbb{E}\left\{P_{k}^{\mathrm{d}}\right\}$ is the average transmit power allocated to the $k$-th user, $P^{\mathrm{c,nt}}$ the circuit power consumption of each antenna, and $P^{\mathrm{c},\mathrm{na}}$ the circuit power consumption of carrier frequency configuration.
	
	Then, the average total power consumed by UL and DL transmissions can be obtained as follows:
	\begin{equation}
		P_{\mathrm {tot}} = \omega^\mathrm{u} \sum_{m \in \mathcal{S}} \mathbb{E}\left\{P_{\mathrm {tot},m}^{\mathrm{u}}\right\}+ \omega^\mathrm{d} \mathbb{E}\left\{P_{\mathrm {tot}}^{\mathrm{d}}\right\},\label{eq:power_av}
	\end{equation}
	where $\omega^\mathrm{u}$ and $\omega^\mathrm{d}$ are weighing factors of UL and DL average total power, respectively. With setting $\omega^\mathrm{u}=\omega^\mathrm{d}=$ 1, the upper bound of $P_{\mathrm {tot}}$ can be obtained by substituting (\ref{eq:ULpower_av}) and (\ref{eq:DLpower_av}) into (\ref{eq:power_av}):
	\begin{equation}
		\begin{aligned}
			P_{\mathrm {tot}}^{\mathrm {UB}}\triangleq &\sum_{m \in \mathcal{S}}\left[\sum_{j=0}^{N_{\mathrm{a}}-1}\left(\varepsilon^{\mathrm{p}, \mathrm{u}}\right)^{\frac{j}{N_{\mathrm{a}}}}\frac{\kappa\phi N_{0} B_{m}^{\mathrm{u}} \gamma_{m}^{\mathrm{th},\mathrm{u}}}{\rho^\mathrm{u}\mu_{m}^{\mathrm{u}} \left(N_\mathrm{t}-1\right)}+P^{\mathrm{c,u}}\right]\\
			&+\sum_{k \in \mathcal{U}}\frac{\xi_{k} \phi N_{0} B_{k}^{\mathrm{d}} \gamma_{k}^{\mathrm{th},\mathrm{d}}}{\rho^\mathrm{d}\mu_{k}^{\mathrm{d}}\left(N_\mathrm{t}-1\right)}+N_\mathrm{t}P^{\mathrm{c,nt}}+\frac{P^{\mathrm{c,na}}}{N_\mathrm{a}}.
		\end{aligned}\label{eq:power_bound}
	\end{equation}
	
	The EE is defined as the ratio of average throughput to average total power consumption. For a given arrival process in a system, the throughput refers to the number of bits actually transmitted per second, rather than the maximum number of bits that the system can transmit per second (i.e., capacity). When the queue is in steady state, the departure rate is equal to the arrival rate, and hence the average throughput is $L\sum_{k\in \mathcal{U}}\lambda_{k}/T_\mathrm{f}$. Then, the EE of URLLC can be defined as follows:		
	\begin{equation}
		\eta_\mathrm{EE}\triangleq \frac{\left(L\sum_{k\in \mathcal{U}}\lambda_{k}/T_\mathrm{f}\right)\left(1-\varepsilon_\mathrm{max}\right)}{P_{\mathrm {tot}}^{\mathrm {UB}}}.\label{eq:EE}
	\end{equation}
	
	\subsection{Optimization Problem}
	The EE is expected to be improved as much as possible by resource allcation. From (\ref{eq:EE}), we can see that reducing power consumption enhances EE since other parameters of throughput are specified. Therefore, maximizing EE is equivalent as minimizing power consumption.
	
	As shown in (\ref{eq:power_bound}), $P_{\mathrm {tot}}^{\mathrm {UB}}$ depends on the allocated bandwidth to each sensor or user in UL or DL, the number of antennas at the BS, as well as the packet dropping probability and the number of assigned subchannels in UL. Although the transmit power thresholds in UL and DL do not directly affect $P_{\mathrm {tot}}^{\mathrm {UB}}$, they control the values of bandwidth and the number of subchannels and antennas. Therefore, the transmit power thresholds indirectly affect $P_{\mathrm {tot}}^{\mathrm {UB}}$. The optimization problem, which minimizes the upper bound of the average total power consumption under the QoS constraints, can be formulated as follows:
	\begin{subequations}
		\begin{align}
			\min_{P_{m}^{\mathrm{th},\mathrm{u}},P_{k}^{\mathrm{th},\mathrm{d}},\atop B_{m}^{\mathrm{u}},B_{k}^{\mathrm{d}},N_{\mathrm{a}},N_{\mathrm{t}}}\ &P_{\mathrm{tot}}^{\mathrm{UB}}\tag{\ref{eq:OP}}\\
			\mbox{s.t.}\ &\sum_{m\in \mathcal{S}} B_{m}^{\mathrm{u}}+\sum_{k\in \mathcal{U}} B_{k}^{\mathrm{d}}\leq W_\mathrm{max}, \tag{\ref{eq:OP}a}\label{OPa}\\ 
			&P_m^{\mathrm{th},\mathrm{u}}\leq P_{\mathrm{max}}^{\mathrm{u}}, \tag{\ref{eq:OP}b}\label{OPb}\\
			&\sum_{k\in\mathcal{U}}P_k^{\mathrm{th},\mathrm{d}}\leq P_{\mathrm{max}}^{\mathrm{d}}, \tag{\ref{eq:OP}c}\label{OPc}\\
			&\varepsilon^{\mathrm{c}, \mathrm{u}}=\varepsilon^{\mathrm{p}, \mathrm{u}}=\varepsilon^{\mathrm{c}, \mathrm{d}}=\varepsilon^{\mathrm{p}, \mathrm{d}}=\varepsilon^{\mathrm{q}}=\frac{\varepsilon_{\mathrm{max}}}{5}, \tag{\ref{eq:OP}d}\label{OPd}\\
			&1\leq N_{\mathrm{a}}\leq N_\mathrm{a}^\mathrm{max},\tag{\ref{eq:OP}e}\label{OPe}\\
			&\textrm{(\ref{eq:UL_SNR})},\textrm{(\ref{eq:ULdropping})},\textrm{(\ref{eq:DL_SNR})},\textrm{(\ref{eq:DLdropping})},\notag \\
			&0<B_{m}^{\mathrm{u}}<W_\mathrm{c},0<B_{k}^{\mathrm{d}}<W_\mathrm{c},\notag\\
			&P_{m}^{\mathrm{th},\mathrm{u}},P_{k}^{\mathrm{th},\mathrm{d}},N_{\mathrm{t}}>0, \notag
		\end{align}\label{eq:OP}
	\end{subequations}
	where (\ref{OPa}) is the maximum bandwidth constraint, (\ref{OPb}) and (\ref{OPc}) are the maximum UL and DL transmit power constraints respectively, (\ref{OPd}) is a near optimal combination of packet loss/error probabilities that ensures the overall packet loss probability in (\ref{eq:packet loss}),\footnote{As disscussed in \cite{She2018twc127}, compared with the optimal combination, setting an equally-divided combination only causes minor performance loss in transmit power. Moreover, the transmit power will go to infinity if any one of these components goes to zero. Therefore, setting a five equal division is a reasonable way to simplify the resource allocation optimization.}(\ref{OPe}) is the assigned subchannel constraint in order that the E2E latency in (\ref{eq:E2E latency}) is satisfied, (\ref{eq:UL_SNR}) and (\ref{eq:DL_SNR}) guarantee the UL decoding error probability as well as the DL decoding error probability and queueing delay violation probability, while (\ref{eq:ULdropping}) and (\ref{eq:DLdropping}) guarantee the UL and DL packet dropping probabilities, respectively.
	
	\section{Joint UL and DL Resource Allocation}\label{V}
\begin{figure}[!t]
		\centering
		\includegraphics[width=3.5in]{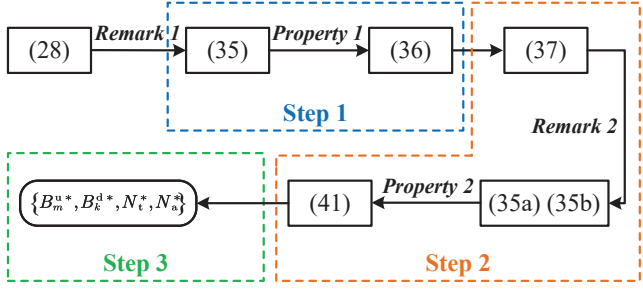}
		\caption{Outline of the derivation in Section \ref{V}.}
		\label{fig:process}
	\end{figure}

	In this section, we first transform problem (\ref{eq:OP}) into an equvialence problem (\ref{eq:OP1}) based on Remark \ref{proposition:1}. Then, we provide a three-step method to find the optimal solution of problem (\ref{eq:OP1}). In the first step, with the arbitrary number of antennas and subchannels, the non-convexity in problem (\ref{eq:OP1}) degenerates into a convex problem (\ref{eq:OP2}) based on Property \ref{property:1}, and then we find the optimal bandwidth allocation. In the second step, we firstly find the required minimum number of antennas by solving another problem (\ref{eq:OP3}), then with given subchannels and corresponding optimal bandwidth allocation,  we find the optimal antenna configuration according to Remark \ref{proposition:2} and Property \ref{property:2}. In the third step, with optimal bandwidth allocation and antenna configuration, we find the optimal subchannel assignment. Finally, we prove that the three-step method can indeed provide the optimal solution. The outline of the derivation is depicted in Fig. \ref{fig:process}, and the process of resource allocation is implemented at a central node. 
	\subsection{Preliminaries for Resource Allocation}
	Based on (\ref{eq:ULdroppingFORMULA}), the solution of $B_{N_\mathrm{t},N_\mathrm{a}}^{\mathrm{u}}\left(g_m^{\mathrm{th},\mathrm{u}}\right)=\varepsilon_{\mathrm{max}}/5$ is identical to each sensor, and thus let $g_m^{\mathrm{th},\mathrm{u}}=g^{\mathrm{th},\mathrm{u}}$. Substitute $g^{\mathrm{th},\mathrm{u}}$ into (\ref{eq:ULgainthreshold}), then the UL transmit power threshold can be obtained as follow:
	\begin{equation}
		P_{m}^{\mathrm{th},\mathrm{u}}= \frac{\phi N_0 B_m^{\mathrm{u}}\gamma_{m}^{\mathrm{th},\mathrm{u}}}{\mu_m^{\mathrm{u}}g^{\mathrm{th},\mathrm{u}}},
		\label{eq:UL_powerthreshold}
	\end{equation}
	which depends on the UL bandwidth allocation ($\gamma_{m}^{\mathrm{th},\mathrm{u}}$ also depends on $B_m^{\mathrm{u}}$). Then, constraint (\ref{OPb}) can be expressed as
	\begin{equation}
		\frac{\phi N_0 B_m^{\mathrm{u}}\gamma_{m}^{\mathrm{th},\mathrm{u}}}{\mu_m^{\mathrm{u}}g^{\mathrm{th},\mathrm{u}}}\leq P_{\mathrm{max}}^{\mathrm{u}}.\label{eq:UL_powerthresholdconstraint}
	\end{equation}
	
	Based on (\ref{eq:DLdroppingFORMULA}), the solution of $B_{N_\mathrm{t},N_\mathrm{a}}^{\mathrm{d}}\left(g_k^{\mathrm{th},\mathrm{d}}\right)=\varepsilon_{\mathrm{max}}/5$ is identical to each user, and thus let $g_k^{\mathrm{th},\mathrm{d}}=g^{\mathrm{th},\mathrm{d}}$. Substitute $g^{\mathrm{th},\mathrm{d}}$ into (\ref{eq:DLgainthreshold}), then the DL transmit power threshold can be obtained as follow:
	\begin{equation}
		P_{k}^{\mathrm{th},\mathrm{d}}= \frac{\phi N_0 B_k^{\mathrm{d}}\gamma_{k}^{\mathrm{th},\mathrm{d}}}{\mu_k^{\mathrm{d}}g^{\mathrm{th},\mathrm{d}}},
		\label{eq:DL_powerthreshold}
	\end{equation}
	which depends on the DL bandwidth allocation ($\gamma_{k}^{\mathrm{th},\mathrm{d}}$ also depends on $B_k^{\mathrm{d}}$). Then, constraint (\ref{OPc}) can be expressed as
	\begin{equation}
		\sum_{k\in\mathcal{U}}\frac{\phi N_0 B_k^{\mathrm{d}}\gamma_{k}^{\mathrm{th},\mathrm{d}}}{\mu_k^{\mathrm{d}}g^{\mathrm{th},\mathrm{d}}}\leq P_{\mathrm{max}}^{\mathrm{d}}.\label{eq:DL_powerthresholdconstraint}
	\end{equation}
	
	To simplify the expression of (\ref{eq:power_bound}), we substitute (\ref{eq:UL_SNR}) and (\ref{eq:DL_SNR}) into $\frac{\phi N_0 B_m^{\mathrm{u}}\gamma_{m}^{\mathrm{th},\mathrm{u}}}{\mu_m^{\mathrm{u}}}$ and $\frac{\phi N_0 B_k^{\mathrm{d}}\gamma_{k}^{\mathrm{th},\mathrm{d}}}{\mu_k^{\mathrm{d}}}$, respectively. Then, two equations are defined as follows:
	\begin{align}
		\Upsilon_{m}^{\mathrm{u}}\left(B_m^{\mathrm{u}}\right)&\triangleq \notag \\
		\frac{\phi N_0 B_m^{\mathrm{u}}}{\mu_m^{\mathrm{u}}}&\left\{\exp\left[\frac{u \ln2}{\tau B_m^{\mathrm{u}}}+\frac{f_Q^{-1}\left(\varepsilon^{\mathrm{c},\mathrm{u}}\right)}{\sqrt{\tau B_m^{\mathrm{u}}}}\right]-1\right\},\label{eq:UL_reSNR}\\
		\Upsilon_{k}^{\mathrm{d}}\left(B_k^{\mathrm{d}}\right)&\triangleq \notag\\
		\frac{\phi N_0 B_k^{\mathrm{d}}}{\mu_k^{\mathrm{d}}}&\left\{\exp\left[\frac{E_{k}^{\mathrm{B}}\left(N_{\mathrm{a}}\right) u\ln2}{\tau B_k^{\mathrm{d}}}+\frac{f_Q^{-1}\left(\varepsilon^{\mathrm{c},\mathrm{d}}\right)}{\sqrt{\tau B_k^{\mathrm{d}}}}\right]-1\right\}\label{eq:DL_reSNR},
	\end{align}
	
	Given a sufficietnly large $N_\mathrm{t}$, the resource allocation can be found from the following problem:
	\begin{subequations}
		\begin{align}
			\min_{B_{m}^{\mathrm{u}},B_{k}^{\mathrm{d}},\atop N_{\mathrm{a}},N_{\mathrm{t}}}\ &\frac{1}{N_\mathrm{t}-1}
			\left[\sum_{m \in \mathcal{S}}\frac{1}{\rho^\mathrm{u}}\sum_{j=0}^{N_{\mathrm{a}}-1}\kappa \left(\varepsilon^{\mathrm{p}, \mathrm{u}}\right)^{\frac{j}{N_{\mathrm{a}}}}  \Upsilon_{m}^{\mathrm{u}}\left(B_m^{\mathrm{u}}\right)\right.\notag\\
			&+\left.\frac{1}{\rho^\mathrm{d}}\sum_{k \in \mathcal{U}}\xi_{k}\Upsilon_{k}^{\mathrm{d}}\left(B_k^{\mathrm{d}}\right)\right]+N_\mathrm{t}P^{\mathrm{c,nt}}+\frac{P^{\mathrm{c,na}}}{N_\mathrm{a}}+P^{\mathrm{c,u}}\tag{\ref{eq:OP1}}\notag\\
			\mbox{s.t.}\ &\frac{\Upsilon_{m}^{\mathrm{u}}\left(B_m^{\mathrm{u}}\right)}{g^{\mathrm{th},\mathrm{u}}}\leq P_{\mathrm{max}}^{\mathrm{u}}, \tag{\ref{eq:OP1}a}\label{OP1a}\\ 
			&\sum_{k\in\mathcal{U}}\frac{\Upsilon_{k}^{\mathrm{d}}\left(B_k^{\mathrm{d}}\right)}{g^{\mathrm{th},\mathrm{d}}}\leq P_{\mathrm{max}}^{\mathrm{d}}, \tag{\ref{eq:OP1}b}\label{OP1b}\\
			&\textrm{(\ref{eq:ULdropping})}, \textrm{(\ref{eq:DLdropping})},\textrm{(\ref{OPa})},\textrm{(\ref{OPd})},\textrm{(\ref{OPe})},\notag\\
			&0<B_{m}^{\mathrm{u}}<W_\mathrm{c},0<B_{k}^{\mathrm{d}}<W_\mathrm{c},N_{\mathrm{t}}>0,\notag
		\end{align}\label{eq:OP1}
	\end{subequations}
	where constraints (\ref{OP1a}) and (\ref{OP1b}) are obtained by substituting (\ref{eq:UL_reSNR}) and (\ref{eq:DL_reSNR}) into (\ref{eq:UL_powerthreshold}) and (\ref{eq:DL_powerthreshold}), respectively.
	
	\begin{prn}
		$N_{\mathrm{t}}$ should not be too small, otherwise problem (\ref{eq:OP1}) is infeasible since constraint (\ref{OPd}) cannot be satisfied. Moreover, if $N_{\mathrm{t}}$ is not sufficiently large, the channel gain thresholds will be reduced to ensure constraint (\ref{OPd}), but this may cause (\ref{OP1a}) and (\ref{OP1b}) unsatisfied.
		\label{proposition:1}
	\end{prn}
	\begin{prf}
		Please see Appendix \ref{appen:A}.\hfill\ensuremath{\square}
	\end{prf}
	
	It is worth noting that some variables and constraints in problem (\ref{eq:OP}) are not included in problem (\ref{eq:OP1}). Variables $P_{m}^{\mathrm{th},\mathrm{u}}$ and $P_{k}^{\mathrm{th},\mathrm{d}}$ are determined by $B_{m}^{\mathrm{u}}$ and $B_{k}^{\mathrm{d}}$. Constraints (\ref{eq:UL_SNR}) and (\ref{eq:DL_SNR}) are used to obtain $\Upsilon_{m}^{\mathrm{u}}\left(B_m^{\mathrm{u}}\right)$ and $\Upsilon_{k}^{\mathrm{d}}\left(B_k^{\mathrm{d}}\right)$. Therefore, the optimal solution of problem (\ref{eq:OP1}) satisfies constraints (\ref{eq:UL_SNR}) and (\ref{eq:ULdropping}), and derives $P_{m}^{\mathrm{th},\mathrm{u}*}$ and $P_{k}^{\mathrm{th},\mathrm{d}*}$ based on (\ref{eq:UL_powerthreshold}) and (\ref{eq:DL_powerthreshold}).
	
	However, $\Upsilon_{m}^{\mathrm{u}}\left(B_m^{\mathrm{u}}\right)$ and $\Upsilon_{k}^{\mathrm{d}}\left(B_k^{\mathrm{d}}\right)$ in problem (\ref{eq:OP1}) are non-convex with respect to bandwidth. Fortunately, by applying the theoretical analysis in \cite{She2019twc402}, $\Upsilon_{m}^{\mathrm{u}}\left(B_m^{\mathrm{u}}\right)$ and $\Upsilon_{k}^{\mathrm{d}}\left(B_k^{\mathrm{d}}\right)$ can still preserve the monotonicity and convexity.
	\begin{pry}
		As proven in \cite{She2019twc402}, $\Upsilon_{m}^{\mathrm{u}}\left(B_m^{\mathrm{u}}\right)$ decreases with $B_m^{\mathrm{u}}$ when $0<B_m^{\mathrm{u}}<W_m^{\mathrm{th},\mathrm{u}}$, and increases with $B_m^{\mathrm{u}}$ when $B_m^{\mathrm{u}}>W_m^{\mathrm{th},\mathrm{u}}$. Therefore, $\Upsilon_{m}^{\mathrm{u}}\left(B_m^{\mathrm{u}}\right)$ is minimized at $W_m^{\mathrm{th},\mathrm{u}}$. Moreover, $\Upsilon_{m}^{\mathrm{u}}\left(B_m^{\mathrm{u}}\right)$ is srtictly convex in $B_m^{\mathrm{u}}$ when $0<B_m^{\mathrm{u}}\leq W_m^{\mathrm{th},\mathrm{u}}$. For $\Upsilon_{k}^{\mathrm{d}}\left(B_k^{\mathrm{d}}\right)$, there also exists a unique solution $W_k^{\mathrm{th},\mathrm{d}}$ which is applicable for the above-mentioned corresponding features.\label{property:1}
	\end{pry}
	
	\subsection{Resource Allocation via the Three-Step Method}
	\subsubsection{Step 1}
	In this step, we fix the values of $N_\mathrm{t}$ and $N_\mathrm{a}$, and optimize the values of $P_{m}^{\mathrm{th},\mathrm{u}}$, $P_{k}^{\mathrm{th},\mathrm{d}}$, $B_m^{\mathrm{u}}$ and $B_k^{\mathrm{d}}$. 
	
	Let $g^{\mathrm{th},\mathrm{u}*}$ be the solution of $B_{N_\mathrm{t},N_\mathrm{a}}^{\mathrm{u}}(g_m^{\mathrm{th},\mathrm{u}})=\varepsilon_{\mathrm{max}}/5$, and $g^{\mathrm{th},\mathrm{d}*}$ be the solution of $B_{N_\mathrm{t},N_\mathrm{a}}^{\mathrm{d}}(g_k^{\mathrm{th},\mathrm{d}})=\varepsilon_{\mathrm{max}}/5$. Define the feasible solutions of problem (\ref{eq:OP1}) as $\widetilde{\bold{B}}^{\mathrm{u}}\triangleq [\widetilde{B}_1^{\mathrm{u}},\ldots,\widetilde{B}_{\left|\mathcal{S}\right|}^{\mathrm{u}}]$ and $\widetilde{\bold{B}}^{\mathrm{d}}\triangleq [\widetilde{B}_1^{\mathrm{d}},\ldots,\widetilde{B}_{\left|\mathcal{U}\right|}^{\mathrm{d}}]$. The optimal solutions are denoted as $B_m^{\mathrm{u}*}$ and $B_k^{\mathrm{d}*}$. According to Property \ref{property:1}, if $\widetilde{B}_m^{\mathrm{u}}<W_m^{\mathrm{th},\mathrm{u}}$, then $\Upsilon_{m}^{\mathrm{u}}(\widetilde{B}_m^{\mathrm{u}})>\Upsilon_{m}^{\mathrm{u}}(W_m^{\mathrm{th},\mathrm{u}})$, and hence $B_m^{\mathrm{u}*}<W_m^{\mathrm{th},\mathrm{u}}$. For the case when $\widetilde{B}_m^{\mathrm{u}}>W_m^{\mathrm{th},\mathrm{u}}$, we can construct another feasible solution $\widetilde{B}_m^{\mathrm{u}(a)}$ that satisifies $\widetilde{B}_m^{\mathrm{u}(a)}<W_m^{\mathrm{th},\mathrm{u}}$ and $\Upsilon_{m}^{\mathrm{u}}(\widetilde{B}_m^{\mathrm{u}(a)})\leq\Upsilon_{m}^{\mathrm{u}}(\widetilde{B}_m^{\mathrm{u}})$, such that $B_m^{\mathrm{u}*}$ can be obtained in the convex region $0<B_m^{\mathrm{u}}\leq W_m^{\mathrm{th},\mathrm{u}}$. The above analysis is also applicable for $B_k^{\mathrm{d}*}$ of $\Upsilon_{k}^{\mathrm{d}}\left(B_k^{\mathrm{d}}\right)$. Therefore, the optimal bandwidth allocation can be obtained by solving the following convex problem:\footnote{The non-negative weighted sum here preserves the convexity of functions.}
	\begin{subequations}
		\begin{align}
			\min_{B_{m}^{\mathrm{u}},B_{k}^{\mathrm{d}}}\ &\frac{1}{N_\mathrm{t}-1}
			\left[\sum_{m \in \mathcal{S}}\frac{1}{\rho^\mathrm{u}}\sum_{j=0}^{N_{\mathrm{a}}-1}\kappa \left(\varepsilon^{\mathrm{p}, \mathrm{u}}\right)^{\frac{j}{N_{\mathrm{a}}}}  \Upsilon_{m}^{\mathrm{u}}\left(B_m^{\mathrm{u}}\right)\right.\notag\\
			&+\left.\frac{1}{\rho^\mathrm{d}}\sum_{k \in \mathcal{U}}\xi_{k}\Upsilon_{k}^{\mathrm{d}}\left(B_k^{\mathrm{d}}\right)\right]\tag{\ref{eq:OP2}}\notag\\
			\mbox{s.t.}\ &\textrm{(\ref{OP1a})},\textrm{(\ref{OP1b})},\textrm{(\ref{OPa})},\notag\\
			&0<B_{m}^{\mathrm{u}}<\min \{W_\mathrm{c},W_m^{\mathrm{th},\mathrm{u}}\},\notag\\
			&0<B_{k}^{\mathrm{d}}<\min \{W_\mathrm{c},W_k^{\mathrm{th},\mathrm{d}}\}.\notag
		\end{align}\label{eq:OP2}
	\end{subequations}
	Subsequently, $P_{m}^{\mathrm{th},\mathrm{u}*}$ is obtained by substituting $B_m^{\mathrm{u}*}$ and $g^{\mathrm{th},\mathrm{u}*}$ into (\ref{eq:UL_powerthreshold}). Similarly, $P_{k}^{\mathrm{th},\mathrm{d}*}$ is obtained by substituting $B_k^{\mathrm{d}*}$ and $g^{\mathrm{th},\mathrm{d}*}$ into (\ref{eq:DL_powerthreshold}).
	
	\begin{algorithm}[!t]
		\renewcommand{\algorithmicrequire}{\textbf{Input:}}
		\renewcommand{\algorithmicensure}{\textbf{Output:}}
		\caption{Finding $N_{\mathrm{t}}^{\mathrm{min}}$ and $N_{\mathrm{t}(N_\mathrm{a})}^{\mathrm{in}}$ via binary search} 
		\label{alg1}
		\begin{algorithmic}[1]
			\REQUIRE $\mu$, $T_f$, $D_\mathrm{max}$, $\varepsilon_\mathrm{max}$, $P_{\mathrm{max}}^{\mathrm{u}}$, $P_{\mathrm{max}}^{\mathrm{d}}$, $W_{\mathrm{max}}$, $N_\mathrm{a}^\mathrm{max}$, $\Psi$.
			\ENSURE $N_{\mathrm{t}}^{\mathrm{min}}$, $N_{\mathrm{t}(N_\mathrm{a})}^{\mathrm{in}}$.
			\FOR{$N_\mathrm{a}=1$ \textbf{to} $N_\mathrm{a}^\mathrm{max}$}
			\STATE Set $N_\mathrm{t}^\mathrm{lb}=0$, $N_\mathrm{t}^\mathrm{ub}=\Psi$, $N_\mathrm{t}^\mathrm{bs}=\lceil 0.5( N_\mathrm{t}^\mathrm{lb}+N_\mathrm{t}^\mathrm{ub})\rceil$.
			\WHILE{$N_\mathrm{t}^\mathrm{ub}-N_\mathrm{t}^\mathrm{lb}>1$ \& $N_\mathrm{t}^\mathrm{bs}\geq 2$}
			\STATE Obtain $z^*$ by solving convex problem (\ref{eq:OP3}).
			\IF{$z^*\leq 0$}
			\STATE $N_{\mathrm{t}(N_\mathrm{a})}^{\mathrm{min}}=N_\mathrm{t}^\mathrm{bs}$.
			\STATE $N_\mathrm{t}^\mathrm{ub}=N_\mathrm{t}^\mathrm{bs}$, $N_\mathrm{t}^\mathrm{bs}=\lceil 0.5( N_\mathrm{t}^\mathrm{lb}+N_\mathrm{t}^\mathrm{ub})\rceil$.
			\ELSE
			\STATE $N_\mathrm{t}^\mathrm{lb}=N_\mathrm{t}^\mathrm{bs}$, $N_\mathrm{t}^\mathrm{bs}=\lceil 0.5( N_\mathrm{t}^\mathrm{lb}+N_\mathrm{t}^\mathrm{ub})\rceil$.
			\ENDIF
			\ENDWHILE
			\ENDFOR
			\STATE $N_{\mathrm{t}}^{\mathrm{min}}=\max \left\{N_{\mathrm{t}(N_\mathrm{a})}^{\mathrm{min}}\right\}$.		
			\FOR{$N_\mathrm{a}=1$ \textbf{to} $N_\mathrm{a}^\mathrm{max}$}
			\STATE Set $N_\mathrm{t}^\mathrm{lb}=N_{\mathrm{t}}^{\mathrm{min}}$, $N_\mathrm{t}^\mathrm{ub}=\Psi$, $N_\mathrm{t}^\mathrm{bs}=\lceil 0.5( N_\mathrm{t}^\mathrm{lb}+N_\mathrm{t}^\mathrm{ub})\rceil$.
			\WHILE{$N_\mathrm{t}^\mathrm{ub}-N_\mathrm{t}^\mathrm{lb}>1$ \& $N_\mathrm{t}^\mathrm{bs}\geq 2$}
			\STATE Find $B_m^{\mathrm{u}*}$ and $B_k^{\mathrm{d}*}$ by solving problem (\ref{eq:OP2}).
			\IF{$z\left(B_m^{\mathrm{u}*},B_k^{\mathrm{d}*}\right)<0$}
			\STATE $N_{\mathrm{t}(N_\mathrm{a})}^{\mathrm{in}}=N_\mathrm{t}^\mathrm{bs}$.
			\STATE $N_\mathrm{t}^\mathrm{ub}=N_\mathrm{t}^\mathrm{bs}$, $N_\mathrm{t}^\mathrm{bs}=\lceil 0.5( N_\mathrm{t}^\mathrm{lb}+N_\mathrm{t}^\mathrm{ub})\rceil$.	
			\ELSIF{$z\left(B_m^{\mathrm{u}*},B_k^{\mathrm{d}*}\right)=0$}
			\STATE $N_{\mathrm{t}(N_\mathrm{a})}^{\mathrm{in}}=N_\mathrm{t}^\mathrm{ub}$.
			\STATE \textbf{break while}
			\ENDIF
			\ENDWHILE
			\ENDFOR
		\end{algorithmic}
	\end{algorithm}
	\subsubsection{Step 2}
	In this step, we firstly find the required minimum number of antennas that satisfies constraints (\ref{OP1a}) and (\ref{OP1b}), and then we optimize the value of $N_\mathrm{t}$ with fixed $N_\mathrm{a}$.
	
	As discussed in Remark \ref{proposition:1}, when $N_{\mathrm{t}}$ is not so large, problem (\ref{eq:OP2}) may be infeasible due to (\ref{OP1a}) and (\ref{OP1b}) being violated. Then, the feasibility of problem (\ref{eq:OP2}) can be verified by solving the following convex problem:\footnote{Pointwise maxization preserves the convexity of functions.}
	\begin{subequations}
		\begin{align}
			\min_{B_{m}^{\mathrm{u}},B_{k}^{\mathrm{d}}}\ &z=\max \left\{\max _{m \in \mathcal{S}}\left\{\frac{\Upsilon_{m}^{\mathrm{u}}\left(B_m^{\mathrm{u}}\right)}{g^{\mathrm{th},\mathrm{u}*}}\right\}-P_{\mathrm{max}}^{\mathrm{u}},\right.\notag\\	&\left.\sum_{k\in\mathcal{U}}\frac{\Upsilon_{k}^{\mathrm{d}}\left(B_k^{\mathrm{d}}\right)}{g^{\mathrm{th},\mathrm{d}*}}-P_{\mathrm{max}}^{\mathrm{d}}\right\}\tag{\ref{eq:OP3}}\\
			\mbox{s.t.}\ &\textrm{(\ref{OPa})},0<B_{m}^{\mathrm{u}}<\min \{W_\mathrm{c},W_m^{\mathrm{th},\mathrm{u}}\},\notag \\
			&0<B_{k}^{\mathrm{d}}<\min \{W_\mathrm{c},W_k^{\mathrm{th},\mathrm{d}}\}.\notag
		\end{align}\label{eq:OP3}
	\end{subequations}
	Let $z^*$ denote the minimum value of $z$. Problem (\ref{eq:OP2}) is feasible if and only if $z^*\leq 0$.
	
	The pair of values that corresponds to $z^*\leq 0$ is denoted as $(N_{\mathrm{t}}, N_{\mathrm{a}})$. Let $N_{\mathrm{t}(N_\mathrm{a})}^{\mathrm{min}}$ denote the minimum value of $N_{\mathrm{t}}$ for given $N_{\mathrm{a}}$ that makes problem (\ref{eq:OP2}) feasible. The upper bound of all possible values of $N_{\mathrm{t}(N_\mathrm{a})}^{\mathrm{min}}$ is defined as $N_{\mathrm{t}}^{\mathrm{min}}\triangleq \max\{N_{\mathrm{t}(N_\mathrm{a})}^{\mathrm{min}}\}$. When $N_{\mathrm{t}}\geq N_{\mathrm{t}}^{\mathrm{min}}$, $P_{\mathrm{tot}}^{\mathrm{UB}}$ is obatined as follows:
	\begin{equation}
		P_{\mathrm{tot}}^{\mathrm{UB}}=\frac{\Omega\left(N_\mathrm{t},N_\mathrm{a}\right)}{N_\mathrm{t}-1}+N_\mathrm{t}P^{\mathrm{c,nt}}+\frac{P^{\mathrm{c,na}}}{N_\mathrm{a}}+P^{\mathrm{c,u}},\label{eq:power_bound1}
	\end{equation}
	where
	\begin{equation}
		\begin{aligned}
			&\Omega\left(N_\mathrm{t},N_\mathrm{a}\right)\\
			\triangleq&\sum_{m \in \mathcal{S}}\frac{1}{\rho^\mathrm{u}}\sum_{j=0}^{N_{\mathrm{a}}-1}\kappa \left(\varepsilon^{\mathrm{p}, \mathrm{u}}\right)^{\frac{j}{N_{\mathrm{a}}}}  \Upsilon_{m}^{\mathrm{u}}\left(B_m^{\mathrm{u}*}\right)+\frac{1}{\rho^\mathrm{d}}\sum_{k \in \mathcal{U}}\xi_{k}\Upsilon_{k}^{\mathrm{d}}\left(B_k^{\mathrm{d}*}\right).
		\end{aligned}\label{eq:OMEGA}
	\end{equation}
	Then, the optimal antenna configuration can be found from the following problem:
	\begin{subequations}
		\begin{align}
			\min_{N_\mathrm{t}}\ &\frac{\Omega\left(N_\mathrm{t},N_\mathrm{a}\right)}{N_\mathrm{t}-1}+N_\mathrm{t}P^{\mathrm{c,nt}}+\frac{P^{\mathrm{c,na}}}{N_\mathrm{a}}+P^{\mathrm{c,u}}\tag{\ref{eq:OP4}}\\
			\mbox{s.t.}\ &N_{\mathrm{t}}\geq N_{\mathrm{t}}^{\mathrm{min}}.\notag
		\end{align}\label{eq:OP4}
	\end{subequations}
	
	The optimal solutions of problem (\ref{eq:OP4}) are denoted as $N_{\mathrm{t}}^*$. Intuitively, $B_m^{\mathrm{u}*}$ and $B_k^{\mathrm{d}*}$ depend on $N_{\mathrm{a}}$ and $N_{\mathrm{t}}$. In this way, $B_m^{\mathrm{u}*}$ and $B_k^{\mathrm{d}*}$ has to change with every round optimization of $N_{\mathrm{t}}$. To simply the procedure of finding $N_{\mathrm{t}}^*$, we have the following remark.
	\begin{prn}
		Whether $B_m^{\mathrm{u}*}$ and $B_k^{\mathrm{d}*}$ changing with $N_{\mathrm{t}}$ and $N_{\mathrm{a}}$ depends on whether constraints (\ref{OP1a}) and (\ref{OP1b}) are active. Particularly, $B_m^{\mathrm{u}*}$ and $B_k^{\mathrm{d}*}$ are independent of $N_{\mathrm{t}}$ and $N_{\mathrm{a}}$ in the case where $N_\mathrm{t}$ is large enough.
		\label{proposition:2}
	\end{prn}
	\begin{prf}
		Please see Appendix \ref{appen:B}.\hfill\ensuremath{\square}
	\end{prf}
	
	Let $N_{\mathrm{t}(N_\mathrm{a})}^{\mathrm{in}}$ denote the minimal value of $N_{\mathrm{t}}$ for given $N_{\mathrm{a}}$ that makes constraints (\ref{OP1a}) and (\ref{OP1b}) inactive, which holds $N_{\mathrm{t}}^{\mathrm{min}}<N_{\mathrm{t}(N_\mathrm{a})}^{\mathrm{in}}$. By solving problem (\ref{eq:OP3}) and (\ref{eq:OP2}), $N_{\mathrm{t}}^{\mathrm{min}}$ and $N_{\mathrm{t}(N_\mathrm{a})}^{\mathrm{in}}$ can be found via binary search. The details of the searching are provided in Algorithm \ref{alg1}, where $\Psi$ is a large number. 
	\begin{pry}
		For the case where $N_{\mathrm{t}}^{\mathrm{min}}<N_{\mathrm{t}(N_\mathrm{a})}^{\mathrm{in}}<N_{\mathrm{t}}$, $P_{\mathrm{tot}}^{\mathrm{UB}}$ with the constant (\ref{eq:OMEGA}) denoted as $\Omega$ is minimized at $1+\sqrt{\frac{\Omega}{P^\mathrm{c,nt}}}$.\label{property:2}
	\end{pry}
	\begin{prf}
		Please see Appendix \ref{appen:C}.\hfill\ensuremath{\square}
	\end{prf}
	
	We denote $N_{\mathrm{t}}^{*}$ for given $N_{\mathrm{a}}$ as $N_{\mathrm{t}(N_{\mathrm{a}})}^*$. For the case where $N_{\mathrm{t}}^{\mathrm{min}}<N_{\mathrm{t}(N_\mathrm{a})}^{\mathrm{in}}<N_\mathrm{t}$, (\ref{eq:OMEGA}) is a constant, and hence from Property \ref{property:2}, $N_{\mathrm{t}(N_{\mathrm{a}})}^*$ can be obtained as,
	\begin{align}
		N_{\mathrm{t}(N_{\mathrm{a}})}^*=
		\begin{cases}
			\left\lceil1+\sqrt{\frac{\Omega}{P^\mathrm{c,nt}}}\right\rceil &\text{if } N_{\mathrm{t}(N_\mathrm{a})}^{\mathrm{in}}\leq \left\lceil1+\sqrt{\frac{\Omega}{P^\mathrm{c,nt}}}\right\rceil<N_{\mathrm{t}},\\
			N_{\mathrm{t}(N_\mathrm{a})}^{\mathrm{in}} &\text{if } \left\lceil1+\sqrt{\frac{\Omega}{P^\mathrm{c,nt}}}\right\rceil<N_{\mathrm{t}(N_\mathrm{a})}^{\mathrm{in}}<N_{\mathrm{t}},
		\end{cases}
	\end{align}
	where $\lceil x\rceil$ is the minimal integer not less than $x$. For the case where $N_{\mathrm{t}}^{\mathrm{min}}<N_\mathrm{t}\leq N_{\mathrm{t}(N_\mathrm{a})}^{\mathrm{in}}$, (\ref{eq:OMEGA}) is a variable of $N_{\mathrm{t}}$ and $N_{\mathrm{a}}$, hence $N_{\mathrm{t}(N_{\mathrm{a}})}^*$ can be found via exhaustive search in $[N_{\mathrm{t},}^{\mathrm{min}},N_{\mathrm{t}(N_\mathrm{a})}^{\mathrm{in}}]$. 
	\begin{algorithm}[!t]
		\renewcommand{\algorithmicrequire}{\textbf{Input:}}
		\renewcommand{\algorithmicensure}{\textbf{Output:}}
		\caption{Resource allocation by the three-step method}
		\label{alg2}
		\begin{algorithmic}[1]
			\REQUIRE $\mu$, $T_f$, $D_\mathrm{max}$, $\varepsilon_\mathrm{max}$, $P_{\mathrm{max}}^{\mathrm{u}}$, $P_{\mathrm{max}}^{\mathrm{d}}$, $W_{\mathrm{max}}$, $N_\mathrm{a}^\mathrm{max}$, $\Psi$.
			\ENSURE $B_m^{\mathrm{u}*}$, $B_k^{\mathrm{d}*}$, $N_\mathrm{t}^*$, $N_\mathrm{a}^*$
			\STATE Obtain $N_\mathrm{t}^\mathrm{min}$ and $N_{\mathrm{t}(N_\mathrm{a})}^\mathrm{in}$ with Algorithm \ref{alg1}.
			\FOR{$N_\mathrm{a}=1$ \textbf{to} $N_\mathrm{a}^\mathrm{max}$}
			\STATE Set $N_\mathrm{t}=N_\mathrm{t}^\mathrm{min}$, $i=1$.
			\WHILE{$N_\mathrm{t}\leq N_{\mathrm{t}(N_\mathrm{a})}^\mathrm{in}$}
			\STATE Find $B_{m}^{\mathrm{u}*}$ and $B_{k}^{\mathrm{d}*}$ by solving problem (\ref{eq:OP2}).
			\STATE $B_{m}^{\mathrm{u}(i)}=B_{m}^{\mathrm{u}*}$, $B_{k}^{\mathrm{d}(i)}=B_{k}^{\mathrm{d}*}$, $N_\mathrm{t}^{(i)}=N_\mathrm{t}$.
			\STATE $P_{\mathrm{tot}}^{\mathrm{UB}(i)}=P_{\mathrm{tot}}^{\mathrm{UB}}\left(B_{m}^{\mathrm{u}*},B_{k}^{\mathrm{d}*},N_{\mathrm{t}},N_{\mathrm{a}}\right)$.
			\STATE $N_\mathrm{t}=N_\mathrm{t}+1$, $i=i+1$.
			\ENDWHILE
			\STATE $B_{m}^{\mathrm{u}(i)}=B_{m}^{\mathrm{u}*}$, $B_{k}^{\mathrm{d}(i)}=B_{k}^{\mathrm{d}*}$, $N_\mathrm{t}^{(i)}=N_\mathrm{t}$.
			\STATE Set $\Omega=\Omega\left(N_\mathrm{t},N_\mathrm{a}\right)$, $N_\mathrm{t}^\mathrm{th}=\left\lceil1+\sqrt{\frac{\Omega}{P^\mathrm{c,nt}}}\right\rceil$.
			\IF{$N_{\mathrm{t}(N_\mathrm{a})}^\mathrm{in}\leq N_\mathrm{t}^\mathrm{th}$}
			\STATE $P_{\mathrm{tot}}^{\mathrm{UB}(i)}=P_{\mathrm{tot}}^{\mathrm{UB}}\left(B_{m}^{\mathrm{u}*},B_{k}^{\mathrm{d}*},N_\mathrm{t}^\mathrm{th},N_{\mathrm{a}}\right)$.
			\ELSIF{$N_\mathrm{t}^\mathrm{th}<N_{\mathrm{t}(N_\mathrm{a})}^\mathrm{in}$}
			\STATE $P_{\mathrm{tot}}^{\mathrm{UB}(i)}=P_{\mathrm{tot}}^{\mathrm{UB}}\left(B_{m}^{\mathrm{u}*},B_{k}^{\mathrm{d}*},N_{\mathrm{t}(N_\mathrm{a})}^\mathrm{in},N_{\mathrm{a}}\right)$.
			\ENDIF
			\STATE $B_{m(N_\mathrm{a})}^{\mathrm{u}*},B_{k(N_\mathrm{a})}^{\mathrm{d}*},N_{\mathrm{t}(N_\mathrm{a})}^*= \underset{B_{m}^{\mathrm{u}(i)},B_{k}^{\mathrm{d}(i)},\atop N_\mathrm{t}^{(i)}}{\arg}\min\left\{P_{\mathrm{tot}}^{\mathrm{UB}(i)}\right\}$.
			\ENDFOR
			\STATE $B_m^{\mathrm{u}*},B_k^{\mathrm{d}*},N_\mathrm{t}^*,N_\mathrm{a}^*=\underset{B_{m(N_\mathrm{a})}^{\mathrm{u}*},B_{k(N_\mathrm{a})}^{\mathrm{d}*},\atop N_{\mathrm{t}(N_\mathrm{a})}^*,N_\mathrm{a}}{\arg}\min \left\{P_{\mathrm{tot}(N_\mathrm{a})}^{\mathrm{UB}}\right\}$.
		\end{algorithmic}
	\end{algorithm}
	\subsubsection{Step 3}
	As the values of $N_{\mathrm{a}}$ are bounded, by comparing $P_{\mathrm{tot}}^{\mathrm{UB}}(B_m^{\mathrm{u}*},B_k^{\mathrm{d}*},N_{\mathrm{t}(N_{\mathrm{a}})}^*,N_{\mathrm{a}})$ for $1\leq N_{\mathrm{a}}\leq N_{\mathrm{a}}^{\mathrm{max}}$, $N_{\mathrm{t}}^*$ and $N_{\mathrm{a}}^*$ can be obtained. 
	
	\subsection{Optimality of the Three-Step Method}
	Given an arbitrary solution of problem (\ref{eq:OP1}) in the case where $1\leq N_\mathrm{a}\leq N_\mathrm{a}^\mathrm{max}$ and $N_{\mathrm{t}}\geq N_{\mathrm{t}}^{\mathrm{min}}$: $\widetilde{B}_m^{\mathrm{u}}$, $\widetilde{B}_k^{\mathrm{d}}$, $\widetilde{N}_{\mathrm{t}}$ and $\widetilde{N}_{\mathrm{a}}$, from the first step of the three-step method, the optimal bandwidth allocation is $P_{\mathrm{tot}}^{\mathrm{UB}}(B_m^{\mathrm{u}*},B_k^{\mathrm{d}*},\widetilde{N}_{\mathrm{t}},\widetilde{N}_{\mathrm{a}})$, i.e.,
	\begin{equation}
		P_{\mathrm{tot}}^{\mathrm{UB}}(B_m^{\mathrm{u}*},B_k^{\mathrm{d}*},\widetilde{N}_{\mathrm{t}},\widetilde{N}_{\mathrm{a}})<P_{\mathrm{tot}}^{\mathrm{UB}}(\widetilde{B}_m^{\mathrm{u}},\widetilde{B}_k^{\mathrm{d}},\widetilde{N}_{\mathrm{t}},\widetilde{N}_{\mathrm{a}}).
		\label{eq:first}
	\end{equation}
	According to the second step, the optimal antennas and subchannels for fixed $\widetilde{N}_{\mathrm{a}}$ that minimizes $P_{\mathrm{tot}}^{\mathrm{UB}}(B_m^{\mathrm{u}*},B_k^{\mathrm{d}*},\widetilde{N}_{\mathrm{t}},\widetilde{N}_{\mathrm{a}})$ are $B_m^{\mathrm{u}*}$, $B_k^{\mathrm{d}*}$, $N_{\mathrm{t}}^*$, and hence
	\begin{equation}
		P_{\mathrm{tot}}^{\mathrm{UB}}(B_m^{\mathrm{u}*},B_k^{\mathrm{d}*},N_{\mathrm{t}}^*,\widetilde{N}_{\mathrm{a}})<P_{\mathrm{tot}}^{\mathrm{UB}}(B_m^{\mathrm{u}*},B_k^{\mathrm{d}*},\widetilde{N}_{\mathrm{t}},\widetilde{N}_{\mathrm{a}}).
		\label{eq:second}
	\end{equation}
	In terms of the third step, the ergodic $P_{\mathrm{tot}}^{\mathrm{UB}}(B_m^{\mathrm{u}*},B_k^{\mathrm{d}*},N_{\mathrm{t}}^*,\widetilde{N}_{\mathrm{a}})$ is determined, and hence the optimal resource allocation is $P_{\mathrm{tot}}^{\mathrm{UB}}(B_m^{\mathrm{u}*},B_k^{\mathrm{d}*},N_{\mathrm{t}}^*,N_{\mathrm{a}}^*)$, i.e.,
	\begin{equation}
		P_{\mathrm{tot}}^{\mathrm{UB}}(B_m^{\mathrm{u}*},B_k^{\mathrm{d}*},N_{\mathrm{t}}^*,N_{\mathrm{a}}^*)<P_{\mathrm{tot}}^{\mathrm{UB}}(B_m^{\mathrm{u}*},B_k^{\mathrm{d}*},N_{\mathrm{t}}^*,\widetilde{N}_{\mathrm{a}}).
		\label{eq:third}
	\end{equation}
	
	From (\ref{eq:first}) to (\ref{eq:third}), we have $P_{\mathrm{tot}}^{\mathrm{UB}}(B_m^{\mathrm{u}*},B_k^{\mathrm{d}*},N_{\mathrm{t}}^*,N_{\mathrm{a}}^*)<P_{\mathrm{tot}}^{\mathrm{UB}}(\widetilde{B}_m^{\mathrm{u}},\widetilde{B}_k^{\mathrm{d}},\widetilde{N}_{\mathrm{t}},\widetilde{N}_{\mathrm{a}})$. The details of the three-step method are provided in Algorithm \ref{alg2} where $P_{\mathrm{tot}(N_\mathrm{a})}^{\mathrm{UB}}=\min\left\{P_{\mathrm{tot}}^{\mathrm{UB}(i)}\right\}$.
	
	\subsection{Computational Complexity Analysis}
	The computational complexity of the required antenna finding in each iteration of the binary search in Algorithm \ref{alg1} is dominated by the
	convex problems. Since the optimization problem (\ref{eq:OP3}) consists of $\left|\mathcal{S}\right|+\left|\mathcal{U}\right|$ variables and $\left|\mathcal{S}\right|+\left|\mathcal{U}\right|+1$ constraints, its time complexity is given by $\left(\left|\mathcal{S}\right|+\left|\mathcal{U}\right|\right)\left(\left|\mathcal{S}\right|+\left|\mathcal{U}\right|+1\right)$ (asymptotically $\left(\left|\mathcal{S}\right|+\left|\mathcal{U}\right|\right)^2$) which is polynomial time complexity. Similarly, the optimization problem (\ref{eq:OP2}) consists of $\left|\mathcal{S}\right|+\left|\mathcal{U}\right|$ variables and $2\left|\mathcal{S}\right|+\left|\mathcal{U}\right|+2$ constraints, its time complexity is given by $\left(\left|\mathcal{S}\right|+\left|\mathcal{U}\right|\right)\left(2\left|\mathcal{S}\right|+\left|\mathcal{U}\right|+2\right)$ (asymptotically $\left(\left|\mathcal{S}\right|+\left|\mathcal{U}\right|\right)^2$). Therefore, the overall complexity of Algorithm \ref{alg1} is of order $\mathcal{O}\left(N_\mathrm{a}^\mathrm{max}\left(I_1+I_2\right)\left(\left|\mathcal{S}\right|+\left|\mathcal{U}\right|\right)^2\right)$, where $I_1=\log\left(\Psi+1\right)$ and $I_2=\log\left(\Psi-N_{\mathrm{t}}^{\mathrm{min}}+1\right)$ are the number of iterations required for reaching convergence, respectively.
	
	The computational complexity of the resource allocation in each iteration of the three step method in Algorithm \ref{alg2} is dominated by the antenna finding in Algorithm \ref{alg1} and convex problem (\ref{eq:OP2}). Therefore, the overall complexity of Algorithm \ref{alg2} is of order $\mathcal{O}\left(\left(N_\mathrm{a}^\mathrm{max}\left(I_1+I_2\right)+I_3\right)\left(\left|\mathcal{S}\right|+\left|\mathcal{U}\right|\right)^2\right)$, where $I_3=\sum_{N_\mathrm{a}=1}^{N_\mathrm{a}^\mathrm{max}}N_{\mathrm{t}(N_\mathrm{a})}^\mathrm{in}$.
	
	\section{Simulation Results and Discussion}\label{VI}
	\begin{table}[!t]
		\caption{Parameters\cite{She2018twc127},\cite{She2018tcomm2266},\cite{She2019twc402},\cite{Osseiran2014mcom26} }\label{tab:table1}
		\centering
		\renewcommand\arraystretch{1.45}{
			\setlength{\tabcolsep}{1mm}{
				\begin{tabular}{|l|c|}
					\hline
					Overall packet loss probability $\varepsilon_\mathrm{max}$& $\rm 1\times10^{-7}$\\
					\hline
					E2E latency requirement  $D_\mathrm{max}-D^{\mathrm{b}}$& 1 ms\\
					\hline
					Backhaul delay  $D^{\mathrm{b}}$& 0.1 ms\\
					\hline
					Duration of each frame $T_f$ (equals to TTI)& 0.1 ms\\
					\hline
					Duration of data transmission $\tau$& 0.05 ms\\
					\hline
					Maximum available bandwidth $W_\mathrm{max}$& 100 MHz\\
					\hline
					Coherence bandwidth $W_\mathrm{c}$& 0.5 MHz\\
					\hline
					Packt size $L$& 20 bytes (160 bits)\\
					\hline
					SNR loss coefficient $\phi$& 1.5 (around 2 dB)\\
					\hline
					Path loss model 10lg$\left(\mu\right)$& -35.3-37.6lg$\left(d\right)$ dB\\
					\hline
					Single-sided noise spectral density $N_0$& \makecell[c]{-173 dBm/Hz\\(around $\rm 5\times10^{-15}$ $\mu$W/Hz)}\\
					\hline
					Maximum transmit power of sensor $P_{\mathrm{max}}^{\mathrm{u}}$& 23 dBm (200 mW)\\
					\hline
					Maximum transmit power of BS $P_{\mathrm{max}}^{\mathrm{d}}$& 40 dBm (10 W)\\
					\hline
		\end{tabular}}}
	\end{table}
	In this section, we first validate Remarks \ref{proposition:1} and \ref{proposition:2} with simulation results. Then, we show the impact of resource allocation on the average total power consumption under the proposed packet delivery scheme. Finally, we compare the EE of the proposed scheme with other transmission schemes.
	
	We consider a cellular network that covers the service area of 50 $\sim$ 300 sensors and 10 $\sim$ 100 users, with the radius of 250 m for each cell. The sensors and users are associated to the BSs, and uniformly distributed with distances in [50, 250] m from the BS they associated to. Each user desires the packets from its nearby sensors with their distance less than 50 m, and each sensor is activated to upload packets to the BS with average rate 100 packets/s\cite{Khabazian2013tits380}. According to the results in \cite{She2018twc127}, the queueing delay bound $D_{\mathrm{max}}^{\mathrm{q}}$ should not be too tight to give much pressure on the required SNR. Therefore, we set the allowed maximum number of subchannels $N_\mathrm{a}^\mathrm{max}$ as 6 so that the toughest queueing delay requirement leaves to 0.3 ms. Other parameters are listed in Table \ref{tab:table1}, unless otherwise specified.
	\begin{table*}[!t]
		\caption{Impact of resource allocation on $P_{\mathrm{tot}}^{\mathrm{UB}}$ with the proposed packet delivery mechanism}\label{tab:table2}
		\centering
		\renewcommand\arraystretch{1.45}{
			\setlength{\tabcolsep}{1mm}{
				\begin{tabular}{|l|c|c|c|c|c|c|c|}
					\hline
					Resource allocation strategy & Eq. BW & Fixed $N_{\mathrm{a}}$ & Fixed $N_{\mathrm{t}}$ & Opt. BW & Opt. $N_{\mathrm{a}}$ & Opt. $N_{\mathrm{t}}$ & Opt. BW, $N_{\mathrm{t}}$, $N_{\mathrm{a}}$\\
					\hline
					Average total power consumption (dBm) & 41.97 & 42.08 & 42.22 & 42.52 & 42.43 & 43.14 & 41.69\\
					\hline
					Difference to optimal solution (dB)& 0.28 & 0.39 & 0.53 & 0.83 & 0.74 & 1.45 & -\\
					\hline
					Relative difference to optimal solution & 6.61\%& 9.46\%& 12.89\%& 20.95\%& 18.54\%& 39.75\%& -\\
					\hline
		\end{tabular}}}
	\end{table*}
	\begin{figure}[!t]
		\centering
		\includegraphics[width=3.5in]{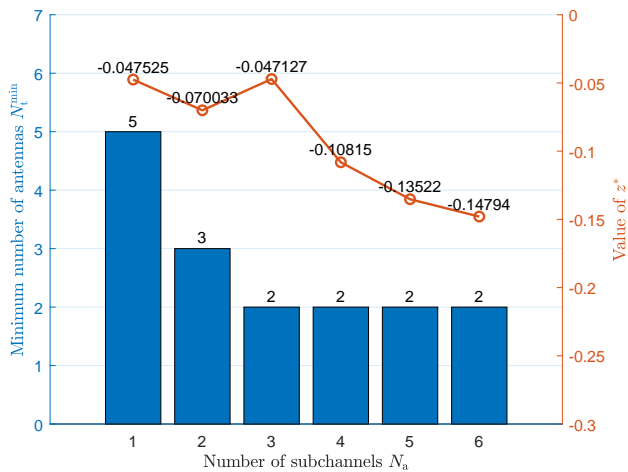}
		\caption{Validation of Remark \ref{proposition:1}, where $\left|\mathcal{S}\right|=200$ and $\left|\mathcal{U}\right|=20$.}
		\label{fig:figure1}
	\end{figure}
	
	Simulation results in Fig. \ref{fig:figure1} validate that the extremely low packet loss probability can be ensured only with sufficient spatial diversity gain, which is quantified as the minimum number of antennas. The simulation is carried out with 200 sensors and 20 users, and $N_\mathrm{t}^\mathrm{min}$ and $z^*$ are obtained according to the optimal solution of problem (\ref{eq:OP3}). For any $N_\mathrm{t}$ less than $N_\mathrm{t}^\mathrm{min}$, the value of $z^*$ is greater than zero, which represents the case where the constraint of required high reliability cannot be satisfied due to the lack of spatial diversity gain. Conversely, $z^*<0$ when $N_\mathrm{t}>N_\mathrm{t}^\mathrm{min}$. This indicates that the resource allocation of problem (\ref{eq:OP}) (equivalently, problem (\ref{eq:OP2}) and (\ref{eq:OP4})) can be solved since the constraints are satisfied. Furthermore, we can see that increasing the number of subchannels may help reduce the required minimum number of antennas up to a point.
	
\begin{figure}[!t]
		\centering
		\includegraphics[width=3.5in]{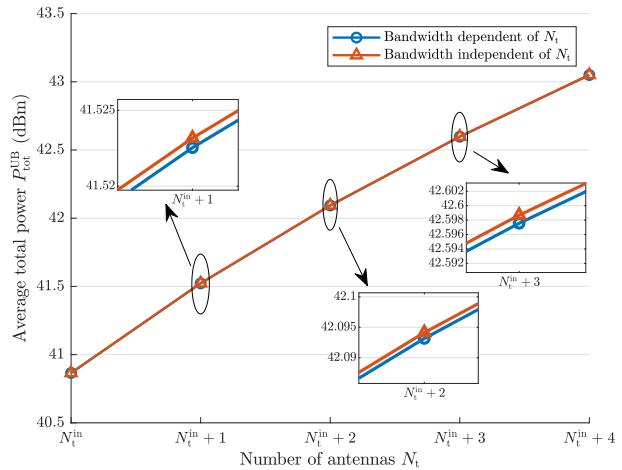}
		\caption{Validation of Remark \ref{proposition:2}, where $\left|\mathcal{S}\right|=200$, $\left|\mathcal{U}\right|=20$ and $N_\mathrm{a}=1$.}
		\label{fig:figure2}
	\end{figure}

	The effect of bandwidth on the value of $P_{\mathrm{tot}}^{\mathrm{UB}}$ in (\ref{eq:power_bound1}) is validated in Fig. \ref{fig:figure2}, for given $N_{\mathrm{t}}^{\mathrm{min}}<N_{\mathrm{t}}^{\mathrm{in}}<N_{\mathrm{t}}$. Resources are allocated according to the optimal solution of problem (\ref{eq:OP2}) and (\ref{eq:OP4}), where $\rho^\mathrm{u}=\rho^\mathrm{d}=$ 0.5, $P^{\mathrm{c,nt}}=$ 33 dBm, $P^{\mathrm{c,na}}=$ 21 dBm, and $P^{\mathrm{c,u}}=$ 18 dBm\cite{Debaillie2015}. The bandwidth dependent of $N_{\mathrm{t}}$ is obtained by solving problem (\ref{eq:OP2}), even if $N_{\mathrm{t}}>N_{\mathrm{t}}^{\mathrm{in}}$. In this case, $P_{\mathrm{tot}}^{\mathrm{UB}}$ (equivalently, $\Omega\left(N_\mathrm{t},N_\mathrm{a}\right)$) changes with the obtained bandwidth in each optimization run. On the contrary, the bandwidth independent of $N_{\mathrm{t}}$ is the solution only when $N_{\mathrm{t}}=N_{\mathrm{t}}^{\mathrm{in}}$, then remains the same when $N_{\mathrm{t}}>N_{\mathrm{t}}^{\mathrm{in}}$. Therefore, $\Omega\left(N_\mathrm{t},N_\mathrm{a}\right)$ stays constant even with the changing $N_{\mathrm{t}}$, which means $P_{\mathrm{tot}}^{\mathrm{UB}}$ is irrelevant to the bandwidth. The gap  of $P_{\mathrm{tot}}^{\mathrm{UB}}$ between the two curves is marginal, and hence corresponds well with the assumption in Remark \ref{proposition:2}. This result also implies that we can omit the optimization of banwidth when $N_{\mathrm{t}}>N_{\mathrm{t}}^{\mathrm{in}}$, so as to simply the process of resource allocation.
	
	\begin{figure}[!t]
		\centering
		\includegraphics[width=3.5in]{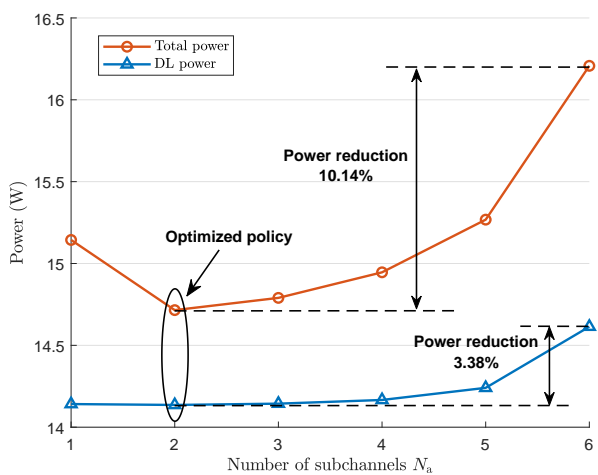}
		\caption{Power consumption vs. number of subchannels, where $\left|\mathcal{S}\right|=300$ and $\left|\mathcal{U}\right|=100$.}
		\label{fig:figure5}
	\end{figure}
	
	Simulation results in Table \ref{tab:table2} show, under the proposed packet delivery mechanism, the impact of resource allocation on $P_{\mathrm{tot}}^{\mathrm{UB}}$ required for ensuring QoS metrics. The results are obtained in the scenario with 300 sensors and 100 users. We compare the optimal resource allocation solution (with header ``Opt. BW, $N_{\mathrm{t}}$, $N_{\mathrm{a}}$'') with other allocation strategies. The header ``Eq. BW'' equally allocates the total bandwidth among sensors and users, whereas the number of antennas and subchannels are optimized. The header ``Fixed $N_\mathrm{a}$'' and ``Fixed $N_\mathrm{t}$'' sets the number of subchannels and antennas as $N_\mathrm{a}^\mathrm{max}$ and $N_\mathrm{t}^\mathrm{in}$, respectively, while the bandwdith and another corresponding variable are optimized (e.g., ``Fixed $N_\mathrm{a}$'' optimizes bandwdith allocation and the number of antennas with $N_\mathrm{a}^\mathrm{max}$ subchannels). The header ``Opt. BW'', ``Opt. $N_\mathrm{a}$'', and ``Opt. $N_\mathrm{t}$'' refer to the single optimization of bandwidth, subchannels, and antennas, while two corresponding variables are fixed (e.g., ``Opt. $N_\mathrm{t}$'' optimizes the number of antennas with equal bandwidth and $N_\mathrm{a}^\mathrm{max}$ subchannles). We can see that the total power obtained via the optimal solution is lower than that via other allocation strategies, and up to nearly 40\% of total power can be saved by optimizing the bandwdith, subchannels, and antennas. Particularly, to show the gain of optimizing the number of subchannels, the power consumption is shown in Fig. \ref{fig:figure5}. Within the allowed number of assigned subchannels, up to around 10\% of the total power can be saved with the optimized number of subchannels. The results also indicate that the required power for DL transmission increases with $N_\mathrm{a}$, which agrees with the higher effective bandwidth due to the increasingly strict queueing delay requirement, leading to a perceptible rise of total power.  
	
	\begin{figure}[!t]
		\centering
		\includegraphics[width=3.5in]{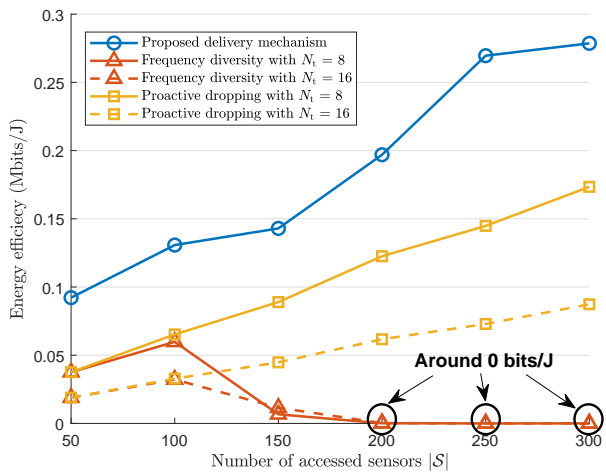}
		\caption{Energy efficiency vs. number of sensors, where $\left|\mathcal{U}\right|=10$.}
		\label{fig:figure3}
	\end{figure}
	
	To show the performace gain of optimal resource allocation under the proposed packet delivery mechanism, we compare the holistic system EE under our scheme with frequency diversity and proactive dropping. Frequency diversity transmits the relipcas of one packet over separated subchannels concurrently, of which each equally assigned subchannel should be less than coherence bandwidth. Therefore, the configuration of separated subchannels is a key factor, whose value is set as in \cite{She2018tcomm2266}. Proactive dropping in \cite{She2018twc127} utilizes the same time-frequency resource to transmit one packet once, hence there is only one subchannel compared with the proposed scheme. 
	
	\begin{figure}[!t]
		\centering
		\includegraphics[width=3.5in]{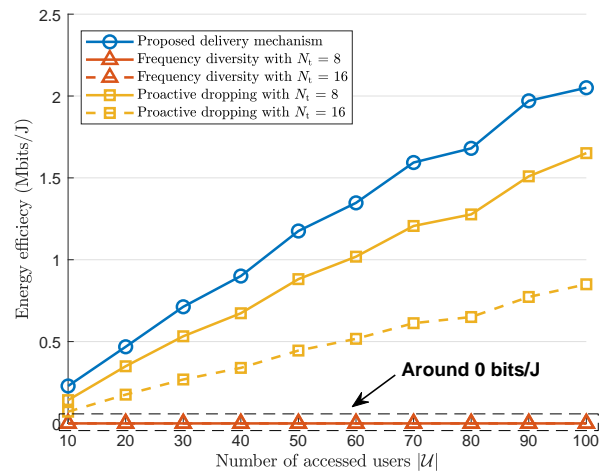}
		\caption{Energy efficiency vs. number of users, where $\left|\mathcal{S}\right|=300$.}
		\label{fig:figure4}
	\end{figure}
	
	For each user to have traffic load from its nearby sensors (i.e., avoid idle state) as much as possible, we respectively set 10 users and 300 sensors when considering the impact of different numbers of sensors and users on EE. The results in Fig. \ref{fig:figure3} and \ref{fig:figure4} show that many more sensors/users can be supported by the proposed delivery mechanism with optimal resource allocation, compared with frequency diveristy (equal bandwidth) and proactive dropping (optimized bandwidth). Particularly, frequency diversity could not support more than 200 devices due to the introduced redundancy via increasing extra bandwidth, which accords with the analysis in Section \ref{III}. Moreover, to further improve the system EE, adjusting the number of antennas according to the number of sensors/users is also an option.
	
	\section{Conclusion}\label{VII}
	In this paper, we have studied joint UL and DL resource allocation to minimize the average total power under the QoS requirements of URLLC and constraint of maximum transmit power, so as to improve the EE. A packet delivery mechanism incorparated with frequency-hopping and proactive dropping was proposed. By providing more chances for transmission and adjusting the packet size at queue buffers, the proposed packet delivery mechanism ensures the target reliability within the target E2E latency even when channel stays in deep fading. To obatin the optimal solution for resource allocation under the proposed mechanism, we have analyzed and validated the remarks required by URLLC transmission, and applied the properties including the convexity with respect to bandwidth, the minimum number of antennas that ensures the feasibility, the independence of bandwidth allocation, and the convexity of antenna configuration with inactive constraints, to the optimization process. Then, a three-step method was proposed, where the bandwidth allocation was firstly optimized with arbitrary antennas and subchannels, then antenna configuration was optimized with given subchannels and results obtained in the first step, eventually we found the optimal subchannels assignment with optimal bandwidth allocation and antenna configuration. Simulation results validated our analysis and showed that the joint optimal resource allocation can save up to nearly 40\% of total power in comparsion with other allocation strageties, and the proposed mechanism not only has higer EE but also supports many more sensors/users in comparsion with other existing transmission schemes.
	
	{\appendices
		\section{Proof of Remark 1}\label{appen:A}
		\setcounter{equation}{0}
		\renewcommand{\theequation}{A.\arabic{equation}}
		\begin{prf}
			As expressed in (\ref{eq:ULdroppingFORMULA}), if $N_{\mathrm{t}}$ goes to infinity, $\varepsilon^{\mathrm{p},\mathrm{u}}$ will be zero and not depend on $g^{\mathrm{th}, \mathrm{u}}$, which can be demonstrated as follows:
			\begin{align}
				&\lim_{N_{\mathrm{t}} \to \infty}\sum_{n=0}^{N_\mathrm{t}-1} \frac{\left(g^{\mathrm{th}, \mathrm{u}}\right)^{n}}{n!}=e^{g^{\mathrm{th}, \mathrm{u}}},\label{eq:A1}\\
				&\lim_{N_{\mathrm{t}} \to \infty}B_{N_\mathrm{t},N_\mathrm{a}}^{\mathrm{u}}\left(g^{\mathrm{th}, \mathrm{u}}\right)=0.\label{eq:A2}
			\end{align}
			(\ref{eq:A1}) and (\ref{eq:A2}) indicate that an extremely small $\varepsilon^{\mathrm{p},\mathrm{u}}$ can be statisified in the case where $N_{\mathrm{t}}$ is large enough.
			
			Similarly, $\varepsilon^{\mathrm{p},\mathrm{d}}$ is statisified when $N_{\mathrm{t}}$ is sufficiently large. The expression of (\ref{eq:DLdroppingFORMULA}) is an upper bound of the derivation in \cite{She2018twc127}, whose original formula should be written as follows:
			\begin{equation}
				F_{N_\mathrm{t},N_\mathrm{a}}^{\mathrm{d}}\left(g^{\mathrm{th}, \mathrm{d}}\right) \triangleq  \int_{0}^{g^{\mathrm{th}, \mathrm{d}}}\left[1-\frac{\ln\left(1+\frac{g\gamma_{k}^{\mathrm{th},\mathrm{d}}}{g^{\mathrm{th}, \mathrm{d}}}\right)}{\ln\left(1+\gamma_{k}^{\mathrm{th},\mathrm{d}}\right)}\right] f_{N_\mathrm{t}}(g) \mathrm{d} g.
			\end{equation} 
			According to the squeeze theorem, we have
			\begin{align}
				0<&\frac{g^{N_{\mathrm{t}}-1}}{\left(N_{\mathrm{t}}-1\right) !}<\left(\frac{g}{N_{\mathrm{t}}-1}\right)^{N_{\mathrm{t}}-1},\\ &\lim_{N_{\mathrm{t}} \to \infty}\left(\frac{g}{N_{\mathrm{t}}-1}\right)^{N_{\mathrm{t}}-1}=0,\\
				&\lim_{N_{\mathrm{t}} \to \infty}\frac{g^{N_{\mathrm{t}}-1}}{\left(N_{\mathrm{t}}-1\right) !}=0.
			\end{align}
			Therefore,
			\begin{equation}
				\lim_{N_{\mathrm{t}} \to \infty}F_{N_\mathrm{t},N_\mathrm{a}}^{\mathrm{d}}\left(g^{\mathrm{th}, \mathrm{d}}\right)=0.
			\end{equation}
			From the above analysis, we know that $N_{\mathrm{t}}$ should not be too small so that problem (\ref{eq:OP1}) can stay feasible.
			
			However, when $N_{\mathrm{t}}$ is not large enough, to satisfy constraint (\ref{OPd}), the channel gain thresholds should be reduced. Here is the demonstration. The first order derivatives of $B_{N_\mathrm{t},N_\mathrm{a}}^{\mathrm{u}}\left(g^{\mathrm{th}, \mathrm{u}}\right)$ and $B_{N_\mathrm{t},N_\mathrm{a}}^{\mathrm{d}}\left(g^{\mathrm{th}, \mathrm{d}}\right)$ can be derived as follows:
			\begin{align}
				{B_{N_\mathrm{t},N_\mathrm{a}}^{\mathrm{u}}}^{'}\left(g^{\mathrm{th}, \mathrm{u}}\right)=&\notag \\
				c_1\left(g^{\mathrm{th}, \mathrm{u}}\right)&c_2\left(g^{\mathrm{th}, \mathrm{u}}\right)N_{\mathrm{a}}\left[1-c_1\left(g^{\mathrm{th}, \mathrm{u}}\right)\right]^{N_{\mathrm{a}}-1},\label{eq:A3}\\
				{B_{N_\mathrm{t},N_\mathrm{a}}^{\mathrm{d}}}^{'}\left(g^{\mathrm{th}, \mathrm{d}}\right)=&\notag \\
				c_1\left(g^{\mathrm{th}, \mathrm{d}}\right)&\left[\frac{N_\mathrm{t}}{\left(g^{\mathrm{th}, \mathrm{d}}\right)^2}+c_2\left(g^{\mathrm{th}, \mathrm{d}}\right)\left(\frac{N_\mathrm{t}}{g^{\mathrm{th}, \mathrm{d}}}-1\right)\right],\label{eq:A4}
			\end{align}	
			where $g=\boldsymbol{h}^{H} \boldsymbol{h}\leq N_\mathrm{t}$, $c_1\left(g\right)=e^{-g} \sum_{n=0}^{N_\mathrm{t}-1} \frac{g^n}{n!}>0$ and $c_2\left(g\right)=\sum_{n=1}^{N_\mathrm{t}-1} \frac{ng^{n-1}}{n!}>0$. 
			
			The each term in (\ref{eq:A3}) and (\ref{eq:A4}) is non-negative. Therefore, ${B_{N_\mathrm{t},N_\mathrm{a}}^{\mathrm{u}}}^{'}\left(g^{\mathrm{th}, \mathrm{u}}\right)>0$ and ${B_{N_\mathrm{t},N_\mathrm{a}}^{\mathrm{d}}}^{'}\left(g^{\mathrm{th}, \mathrm{d}}\right)\geq 0$, which means that $B_{N_\mathrm{t},N_\mathrm{a}}^{\mathrm{u}}\left(g^{\mathrm{th}, \mathrm{u}}\right)$ and $B_{N_\mathrm{t},N_\mathrm{a}}^{\mathrm{d}}\left(g^{\mathrm{th}, \mathrm{d}}\right)$ increase with $g^{\mathrm{th}, \mathrm{u}}$ and $g^{\mathrm{th}, \mathrm{d}}$, respectively. In the case where $N_{\mathrm{t}}$ is not large enough, to satisfy an extremely small $\varepsilon^{\mathrm{p},\mathrm{u}}$ and $\varepsilon^{\mathrm{p},\mathrm{d}}$, $g^{\mathrm{th}, \mathrm{u}}$ and $g^{\mathrm{th}, \mathrm{d}}$ should be reduced, making problem (\ref{eq:OP1}) infeasible due to (\ref{OP1a}) and (\ref{OP1b}) being violated.
			This completes the proof.
		\end{prf}
		
		\section{Proof of Remark 2}\label{appen:B}
		\setcounter{equation}{0}
		\renewcommand{\theequation}{B.\arabic{equation}}
		\begin{prf}
			For the case where constraints (\ref{OP1a}) and (\ref{OP1b}) are active, we have
			\begin{align}
				\frac{\Upsilon_{m}^{\mathrm{u}}\left(B_m^{\mathrm{u}*}\right)}{g^{\mathrm{th},\mathrm{u}}}&=P_{\mathrm{max}}^{\mathrm{u}}\\
				B_m^{\mathrm{u}*}&=\varphi_{m}^{\mathrm{u}\ -1}\left(P_{\mathrm{max}}^{\mathrm{u}}g^{\mathrm{th},\mathrm{u}}\right)\notag \\
				&=\varphi_{m}^{\mathrm{u}\ -1}\left[P_{\mathrm{max}}^{\mathrm{u}}B_{N_\mathrm{t},N_\mathrm{a}}^{\mathrm{u}\quad -1}\left(\varepsilon^{\mathrm{p}, \mathrm{u}}\right)\right],
			\end{align}
			\begin{align}
				\frac{\Upsilon_{k}^{\mathrm{d}}\left(B_m^{\mathrm{d}*}\right)}{g^{\mathrm{th},\mathrm{d}}}&=P_{\mathrm{max}}^{\mathrm{d}}\\
				B_k^{\mathrm{d}*}&=\varphi_{k}^{\mathrm{d}\ -1}\left(P_{\mathrm{max}}^{\mathrm{d}}g^{\mathrm{th},\mathrm{d}}\right)\notag \\
				&=\varphi_{k}^{\mathrm{d}\ -1}\left[P_{\mathrm{max}}^{\mathrm{d}}B_{N_\mathrm{t},N_\mathrm{a}}^{\mathrm{d}\quad -1}\left(\varepsilon^{\mathrm{p}, \mathrm{d}}\right)\right].
			\end{align}
			When the equality holds, $B_m^{\mathrm{u}*}$ and $B_k^{\mathrm{d}*}$ change with $N_{\mathrm{t}}$ and $N_{\mathrm{a}}$ since the other constraints of problem (\ref{eq:OP2}) do not depend on $N_{\mathrm{t}}$ and $N_{\mathrm{a}}$. 
			
			For the case where constraints (\ref{OP1a}) and (\ref{OP1b}) are inactive (i.e., the inequality holds), $B_m^{\mathrm{u}*}$ and $B_k^{\mathrm{d}*}$ are independent of $N_{\mathrm{t}}$ and $N_{\mathrm{a}}$. Moreover, as proven in Remark \ref{proposition:1}, constraints (\ref{OP1a}) and (\ref{OP1b}) are inactive since the channel gain thresholds can be ignored in the circumstance of a sufficiently large $N_{\mathrm{t}}$.
			This completes the proof.
		\end{prf}
		
		\section{Proof of Property 2}\label{appen:C}
		\setcounter{equation}{0}
		\renewcommand{\theequation}{C.\arabic{equation}}
		\begin{prf}
			According to Remark \ref{proposition:2}, Eq. (\ref{eq:OMEGA}) is a constant when $N_{\mathrm{t}}^{\mathrm{min}}<N_{\mathrm{t}(N_\mathrm{a})}^{\mathrm{in}}<N_{\mathrm{t}}$. Given $N_\mathrm{a}$, we define (\ref{eq:power_bound1}) as 
			\begin{equation}
				\label{eq:C1}
				f\left(N_\mathrm{t}\right)\triangleq\frac{C_1}{N_\mathrm{t}-1}+N_\mathrm{t}C_2+\frac{C_3}{N_\mathrm{a}}+C_4,
			\end{equation}
			where $C_1=\Omega$, $C_2=P^{\mathrm{c,nt}}$, $C_3=P^{\mathrm{c,na}}$ and $C_4=P^{\mathrm{c,u}}$.
			
			The first order derivatives of $f\left(N_\mathrm{t}\right)$ is then
			\begin{align}
				\label{eq:C2}
				f^{'}\left(N_\mathrm{t}\right)&=\frac{-C_1}{\left(N_\mathrm{t}-1\right)^2}+C_2\\
				&\begin{cases}
					<0, \quad \text{if } 0<N_\mathrm{t}< 1+\sqrt{\frac{C_1}{C_2}},\notag\\
					\geq 0, \quad \text{if } N_\mathrm{t}\geq 1+\sqrt{\frac{C_1}{C_2}}.
				\end{cases}	
			\end{align}
			The second order derivatives of $f\left(N_\mathrm{t}\right)$ can be derived as follows:
			\begin{align}
				\label{eq:C3}
				f^{''}\left(N_\mathrm{t}\right)&=\frac{-2C_1}{\left(N_\mathrm{t}-1\right)^3}\\
				&\begin{cases}
					>0, \quad \text{if } 0<N_\mathrm{t}< 1,\notag\\
					< 0, \quad \text{if } N_\mathrm{t}> 1.
				\end{cases}
			\end{align}
			
			From (\ref{eq:C2}) and (\ref{eq:C3}), we know $f\left(N_\mathrm{t}\right)$ is convex when $N_\mathrm{t}>1$ and minimized at $1+\sqrt{\frac{C_1}{C_2}}$.
			This completes the proof.
		\end{prf}
	}
	
	\bibliographystyle{IEEEtran} 
	\bibliography{manuscript}

\begin{thebibliography}{10}
\providecommand{\url}[1]{#1}
\csname url@samestyle\endcsname
\providecommand{\newblock}{\relax}
\providecommand{\bibinfo}[2]{#2}
\providecommand{\BIBentrySTDinterwordspacing}{\spaceskip=0pt\relax}
\providecommand{\BIBentryALTinterwordstretchfactor}{4}
\providecommand{\BIBentryALTinterwordspacing}{\spaceskip=\fontdimen2\font plus
\BIBentryALTinterwordstretchfactor\fontdimen3\font minus
  \fontdimen4\font\relax}
\providecommand{\BIBforeignlanguage}[2]{{%
\expandafter\ifx\csname l@#1\endcsname\relax
\typeout{** WARNING: IEEEtran.bst: No hyphenation pattern has been}%
\typeout{** loaded for the language `#1'. Using the pattern for}%
\typeout{** the default language instead.}%
\else
\language=\csname l@#1\endcsname
\fi
#2}}
\providecommand{\BIBdecl}{\relax}
\BIBdecl

\bibitem{You2021SciChinaInfSci}
X.~You, C.-X. Wang, J.~Huang, X.~Gao, Z.~Zhang, M.~Wang, Y.~Huang, C.~Zhang,
  Y.~Jiang, J.~Wang \emph{et~al.}, ``Towards {6G} wireless communication
  networks: Vision, enabling technologies, and new paradigm shifts,''
  \emph{Science China Information Sciences}, vol.~64, no.~1, p.
  110301:1–110301:74, 2021.

\bibitem{Bennis2018jproc1834}
M.~Bennis, M.~Debbah, and H.~V. Poor, ``Ultrareliable and low-latency wireless
  communication: Tail, risk, and scale,'' \emph{Proceedings of the IEEE}, vol.
  106, no.~10, pp. 1834--1853, 2018.

\bibitem{TR38.913}
3GPP, ``Study on scenarios and requirements for next generation access
  technologies,'' 3rd Generation Partnership Project, Technical Report (TR)
  38.913, April 2022.

\bibitem{Chen2018mcom119}
H.~Chen, R.~Abbas, P.~Cheng, M.~Shirvanimoghaddam, W.~Hardjawana, W.~Bao,
  Y.~Li, and B.~Vucetic, ``Ultra-reliable low latency cellular networks: Use
  cases, challenges and approaches,'' \emph{IEEE Communications Magazine},
  vol.~56, no.~12, pp. 119--125, 2018.

\bibitem{Capozzi2013surv678}
F.~Capozzi, G.~Piro, L.~Grieco, G.~Boggia, and P.~Camarda, ``Downlink packet
  scheduling in {LTE} cellular networks: Key design issues and a survey,''
  \emph{IEEE Communications Surveys \& Tutorials}, vol.~15, no.~2, pp.
  678--700, 2013.

\bibitem{TR38.802}
3GPP, ``Study on new radio access technology physical layer aspects,'' 3rd
  Generation Partnership Project, Technical Report (TR) 38.802, September 2017.

\bibitem{TR38.912}
3GPP, ``Study on new radio {(NR)} access technology,'' 3rd Generation
  Partnership Project, Technical Report (TR) 38.912, April 2022.

\bibitem{Shannon1948}
C.~E. Shannon, ``A mathematical theory of communication,'' \emph{The Bell
  System Technical Journal}, vol.~27, no.~3, pp. 379--423, 1948.

\bibitem{She2018twc127}
C.~She, C.~Yang, and T.~Q.~S. Quek, ``Cross-layer optimization for
  ultra-reliable and low-latency radio access networks,'' \emph{IEEE
  Transactions on Wireless Communications}, vol.~17, no.~1, pp. 127--141, 2018.

\bibitem{TR36.814}
3GPP, ``Further advancements for {E-UTRA} physical layer aspects,'' 3rd
  Generation Partnership Project, Technical Report (TR) 36.814, March 2017.

\bibitem{Xu2016twc5527}
S.~Xu, T.-H. Chang, S.-C. Lin, C.~Shen, and G.~Zhu, ``Energy-efficient packet
  scheduling with finite blocklength codes: Convexity analysis and efficient
  algorithms,'' \emph{IEEE Transactions on Wireless Communications}, vol.~15,
  no.~8, pp. 5527--5540, 2016.

\bibitem{Polyanskiy2010tit2307}
Y.~Polyanskiy, H.~V. Poor, and S.~Verdu, ``Channel coding rate in the finite
  blocklength regime,'' \emph{IEEE Transactions on Information Theory},
  vol.~56, no.~5, pp. 2307--2359, 2010.

\bibitem{Polyanskiy2014tit4232}
W.~Yang, G.~Durisi, T.~Koch, and Y.~Polyanskiy, ``Quasi-static multiple-antenna
  fading channels at finite blocklength,'' \emph{IEEE Transactions on
  Information Theory}, vol.~60, no.~7, pp. 4232--4265, 2014.

\bibitem{Popovski2016jproc1711}
G.~Durisi, T.~Koch, and P.~Popovski, ``Toward massive, ultrareliable, and
  low-latency wireless communication with short packets,'' \emph{Proceedings of
  the IEEE}, vol. 104, no.~9, pp. 1711--1726, 2016.

\bibitem{She2018tcomm2266}
C.~She, C.~Yang, and T.~Q.~S. Quek, ``Joint uplink and downlink resource
  configuration for ultra-reliable and low-latency communications,'' \emph{IEEE
  Transactions on Communications}, vol.~66, no.~5, pp. 2266--2280, 2018.

\bibitem{She2019twc402}
C.~Sun, C.~She, C.~Yang, T.~Q.~S. Quek, Y.~Li, and B.~Vucetic, ``Optimizing
  resource allocation in the short blocklength regime for ultra-reliable and
  low-latency communications,'' \emph{IEEE Transactions on Wireless
  Communications}, vol.~18, no.~1, pp. 402--415, 2019.

\bibitem{Liu2022lwc}
B.~Liu, P.~Zhu, J.~Li, D.~Wang, and Y.~Wang, ``Energy-efficient optimization
  via joint power and subcarrier allocation for {eMBB} and {URLLC} services,''
  \emph{IEEE Wireless Communications Letters}, pp. 1--1, 2022.

\bibitem{Pedersen2017mwc154}
K.~I. Pedersen, S.~R. Khosravirad, G.~Berardinelli, and F.~Frederiksen,
  ``Rethink hybrid automatic repeat request design for {5G}: Five configurable
  enhancements,'' \emph{IEEE Wireless Communications}, vol.~24, no.~6, pp.
  154--160, 2017.

\bibitem{Elayoubi2019jsac896}
S.~E. Elayoubi, P.~Brown, M.~Deghel, and A.~Galindo-Serrano, ``Radio resource
  allocation and retransmission schemes for {URLLC} over {5G} networks,''
  \emph{IEEE Journal on Selected Areas in Communications}, vol.~37, no.~4, pp.
  896--904, 2019.

\bibitem{She2016GLOCOMW}
C.~She, C.~Yang, and T.~Q.~S. Quek, ``Uplink transmission design with massive
  machine type devices in tactile internet,'' in \emph{2016 IEEE Globecom
  Workshops (GC Wkshps)}, 2016, pp. 1--6.

\bibitem{She2017GLOCOMW}
C.~Sun, C.~She, and C.~Yang, ``Exploiting multi-user diversity for
  ultra-reliable and low-latency communications,'' in \emph{2017 IEEE Globecom
  Workshops (GC Wkshps)}, 2017, pp. 1--6.

\bibitem{Johansson2015ICCW}
N.~A. Johansson, Y.-P.~E. Wang, E.~Eriksson, and M.~Hessler, ``Radio access for
  ultra-reliable and low-latency {5G} communications,'' in \emph{2015 IEEE
  International Conference on Communication Workshop (ICCW)}, 2015, pp.
  1184--1189.

\bibitem{Feng2019mvt94}
D.~Feng, C.~She, K.~Ying, L.~Lai, Z.~Hou, T.~Q.~S. Quek, Y.~Li, and B.~Vucetic,
  ``Toward ultrareliable low-latency communications: Typical scenarios,
  possible solutions, and open issues,'' \emph{IEEE Vehicular Technology
  Magazine}, vol.~14, no.~2, pp. 94--102, 2019.

\bibitem{TS22.368}
3GPP, ``Service requirements for machine-type communications {(MTC)}; stage
  1,'' 3rd Generation Partnership Project, Technical Specification (TS) 22.368,
  April 2022.

\bibitem{She2018tcomm5482}
C.~She, Z.~Chen, C.~Yang, T.~Q.~S. Quek, Y.~Li, and B.~Vucetic, ``Improving
  network availability of ultra-reliable and low-latency communications with
  multi-connectivity,'' \emph{IEEE Transactions on Communications}, vol.~66,
  no.~11, pp. 5482--5496, 2018.

\bibitem{Tse2005}
D.~Tse and P.~Viswanath, \emph{Fundamentals of Wireless Communication}.\hskip
  1em plus 0.5em minus 0.4em\relax Cambridge University Press, 2005.

\bibitem{She2017SciChinaInfSci}
C.~She and C.~Yang, ``Energy efficiency-{QoS} relation and its application in
  wireless networks,'' \emph{Science China Information Sciences}, vol.~47,
  no.~5, pp. 607--619, 2017.

\bibitem{Liu2015tvt2846}
X.~Liu, S.~Han, and C.~Yang, ``Energy-efficient training-assisted transmission
  strategies for closed-loop {MISO} systems,'' \emph{IEEE Transactions on
  Vehicular Technology}, vol.~64, no.~7, pp. 2846--2860, 2015.

\bibitem{Schiessl2015MSWiM}
S.~Schiessl, J.~Gross, and H.~Al-Zubaidy, ``Delay analysis for wireless fading
  channels with finite blocklength channel coding,'' in \emph{2015 ACM
  International Conference on Modeling, Analysis and Simulation of Wireless and
  Mobile Systems (MSWiM)}, November 2015, pp. 13--22.

\bibitem{Paulraj2008}
A.~Paulraj, R.~Nabar, and D.~Gore, \emph{Introduction to Space-Time Wireless
  Communications}.\hskip 1em plus 0.5em minus 0.4em\relax Cambridge University
  Press, 2008, ch.~3, p. 32–85.

\bibitem{Khabazian2013tits380}
M.~Khabazian, S.~Aissa, and M.~Mehmet-Ali, ``Performance modeling of safety
  messages broadcast in vehicular {Ad Hoc} networks,'' \emph{IEEE Transactions
  on Intelligent Transportation Systems}, vol.~14, no.~1, pp. 380--387, 2013.

\bibitem{Chen1995jsac1091}
C.~Chang and J.~Thomas, ``Effective bandwidth in high-speed digital networks,''
  \emph{IEEE Journal on Selected Areas in Communications}, vol.~13, no.~6, pp.
  1091--1100, 1995.

\bibitem{Choudhury1996tcomm203}
G.~Choudhury, D.~Lucantoni, and W.~Whitt, ``Squeezing the most out of {ATM},''
  \emph{IEEE Transactions on Communications}, vol.~44, no.~2, pp. 203--217,
  1996.

\bibitem{TS22.261}
3GPP, ``Service requirements for the {5G} system,'' 3rd Generation Partnership
  Project, Technical Specification (TS) 22.261, June 2022.

\bibitem{Zhang2016tcomm876}
G.~Zhang, T.~Q.~S. Quek, M.~Kountouris, A.~Huang, and H.~Shan, ``Fundamentals
  of heterogeneous backhaul design—analysis and optimization,'' \emph{IEEE
  Transactions on Communications}, vol.~64, no.~2, pp. 876--889, 2016.

\bibitem{Ashraf2015GC}
S.~A. Ashraf, F.~Lindqvist, R.~Baldemair, and B.~Lindoff, ``Control channel
  design trade-offs for ultra-reliable and low-latency communication system,''
  in \emph{2015 IEEE Globecom Workshops (GC Wkshps)}, 2015, pp. 1--6.

\bibitem{Arnold2010}
O.~Arnold, F.~Richter, G.~Fettweis, and O.~Blume, ``Power consumption modeling
  of different base station types in heterogeneous cellular networks,'' in
  \emph{2010 Future Network \& Mobile Summit}, 2010, pp. 1--8.

\bibitem{Debaillie2015}
B.~Debaillie, C.~Desset, and F.~Louagie, ``A flexible and future-proof power
  model for cellular base stations,'' in \emph{2015 IEEE 81st Vehicular
  Technology Conference (VTC Spring)}, 2015, pp. 1--7.

\bibitem{Osseiran2014mcom26}
A.~Osseiran, F.~Boccardi, V.~Braun, K.~Kusume, P.~Marsch, M.~Maternia,
  O.~Queseth, M.~Schellmann, H.~Schotten, H.~Taoka, H.~Tullberg, M.~A.
  Uusitalo, B.~Timus, and M.~Fallgren, ``Scenarios for {5G} mobile and wireless
  communications: the vision of the {METIS} project,'' \emph{IEEE
  Communications Magazine}, vol.~52, no.~5, pp. 26--35, 2014.

\end{thebibliography}
\end{document}